\documentclass{ar-1col-S2O}

\usepackage[comma]{natbib}
\usepackage{url}
\usepackage{xcolor}
\usepackage{ulem}
\usepackage[utf8]{inputenc}
\usepackage{graphicx}

\newcommand{\feh}{[Fe/H]}
\newcommand{\afe}{[$\alpha$/Fe]}

\newcommand{\lir}{$^6$Li/$^7$Li}
\newcommand{\gcc}{gr cm$^{-3}$}

\newcommand{\lia}{$^6$Li}
\newcommand{\lib}{$^7$Li}
\newcommand{\bea}{$^7$Be}
\newcommand{\beb}{$^9$Be}
\newcommand{\ba}{$^{10}$B}
\newcommand{\bb}{$^{11}$B}
\newcommand{\liba}{$^{7}$Li/$^6$Li}
\newcommand{\bba}{$^{11}$B/$^{10}$B}
\newcommand{\ms}{M$_{\odot}$}
\newcommand{\teff}{T$_{\rm eff}$}

\jname{Xxxx. Xxx. Xxx. Xxx.}
\jvol{AA}
\jyear{YYYY}
\doi{10.1146/((please add article doi))}

\begin{document}

\markboth{Charbonnel \& Prantzos}{Lithium as a probe of stellar and galactic physics}

\title{Lithium as a probe of stellar and galactic physics} 

\author{Corinne Charbonnel$^{1,2}$ and Nikos Prantzos$^3$ 
\affil{$^1$Department of Astronomy, University of Geneva, Chemin P\'egasi 51, 1290 Versoix, Switzerland; email: corinne.charbonnel@unige.ch}
\affil{$^2$IRAP, UMR 5277 CNRS and Université de Toulouse, 14 Av. E.Belin, 31400 Toulouse, France}
\affil{$^3$Institut d’Astrophysique de Paris, UMR7095 CNRS, Sorbonne Universit\'{e}, 98bis Bd. Arago, 75104 Paris, France}}

\begin{abstract}
Lithium plays a unique role in astrophysics, as it is a powerful diagnostic for the physics and evolution of low-mass stars, Galactic archaeology, and cosmology. We review the Li observations in stars at different phases of their evolution, the strengths and the limitations of the current theoretical stellar models to explain the Li abundance data, our understanding of the Li sources and of the evolution of Li throughout the Galactic history.
Key takeaways from  the current state of the research in the field are:
\begin{itemize}
\item Stellar evolution models accounting for fundamental transport processes 

of chemical species  and angular momentum hold the promise of 

providing  a common “stellar Li depletion” explanation to the Li

abundance patterns  observed in all Galactic stellar populations, 

including the "dip" and the "plateau(s)".
\item Novae are most probably the main source of Li in the Galaxy, on 

observational (but not yet theoretically established) grounds.
\item Radial migration of stars in the Galactic disk holds the key to 

understand  many aspects of the 
Li evolution in the Milky Way.
\end{itemize}
\end{abstract}

\maketitle

\tableofcontents


\section{INTRODUCTION}

Lithium (Li), the third lightest element in the periodic table,  holds a unique position in astrophysics as it probes the properties of the outer  layers of stars, the baryonic density of the Universe, and the chemical evolution of the Galaxy.   
Despite the relative simplicity of its atomic structure, the abundance and distribution of Li in stellar environments remain enigmatic and  
continue to challenge our understanding of stellar structure, nucleosynthesis, and galactic chemical evolution. Li  is the only element for which whole workshops or conferences are held (e.g. Paris 2012 or Rome 2019) and monographs are written \citep{Martin_2023}. 

The \lib ~(hereafter Li)  abundance is now available for tens of thousands of low-mass stars in different areas of the colour-magnitude diagram, thanks to the many spectroscopic surveys (e.g. Gaia-ESO, GALAH, LAMOST) that are complementing the ESA Gaia mission  to decipher the history of the Milky Way and asteroseismic surveys that probe the interior of stars (CoRoT, {\it{Kepler}}, K2, TESS, and the future PLATO).

Of all Li discoveries, the one of \cite{Spite_1982} had probably the deepest impact in the fields of stellar and galactic astronomy and cosmology.
Old dwarf stars of the Galactic halo display remarkably constant Li abundances which seem independent of their metal 
content (unlike any other metal) and of their properties, such as effective temperature (unlike the younger stars of the Galactic disk). The original interpretation of this pattern as a relic of the primordial nucleosynthesis met with difficulties twenty years later, after the analysis of observations of the Cosmic Microwave Background (CMB) with the WMAP satellite favoured a higher cosmic density than suggested by the ``halo plateau" of Li. A similar Li plateau has been discovered 
in Galactic dwarf stars from accreted  satellites, including Gaia-Enceladus \citep{Molaro2020,2021MNRAS.507...43S}, 
 and three S2-Stream members \citep{2021MNRAS.500..889A}. 
A variety of solutions were explored, including non-standard primordial nucleosynthesis and/or cosmologies as well as exotic particle physics, but these proved unsuccessful \citep[][and references therein]{Fields_2020}. 
Currently, the most likely answer 
appears to be the “stellar Li depletion solution”.  

\begin{marginnote}
\entry{Metal} {An element heavier than Helium. {\it Metallicity} in stellar astronomy is usually expressed as
\feh=log((Fe/H)$_*$/(Fe/H)$_{\odot}$)}
\end{marginnote}

The two Li isotopes,\lia \ and \lib, are the two most fragile stable nuclides after deuterium (D) and $^3$He. It is this property of nuclear fragility that made Li such a powerful tool enabling astronomers to ``see" below the surfaces of stars and to probe  their subphotospheric layers, well before the era of helio- and asteroseismology. 
In the 1950ies, it was realized that the Li abundance in chondrites - meteoritic inclusions in some of the oldest objects in the solar system - is higher than the one of the solar photosphere 
by more than two orders of magnitude. The Sun has severely depleted its original Li content since its birth in a way that is not predicted by classical stellar models. Various features observed in the Li abundance patterns of field stars and open clusters (OC), like the ``Li-dip", also call for physical processes beyond the ``standard" ones of convection and atomic diffusion.  
While the understanding of the  processes that transport angular momentum and chemicals in stars is one of the hottest topic in stellar physics thanks to the wealth of asteroseismic data, $^7$Li remains a unique and sensitive constraint on evolution models for low-mass stars. As of today, no model ``passes" all the Li tests, but recent theoretical and numerical developments hold the promise of a common ``stellar Li depletion solution" to all the Li abundance patterns observed in the different Galactic stellar populations, including the dip and the plateau.  

The production mechanism of Li - as well as of the other light elements  Be, and B - was an enigma to the founders of stellar nucleosynthesis, who coined the name `x-process' for it \citep{B2FH_1957}. It was progressively realized  that the most abundant isotope of lithium, $^7$Li, is the only nuclide  produced in three different astrophysical sites: in the hot early Universe by primordial nucleosynthesis, in Galactic cosmic rays  through spallation and fusion reactions, and in stars by thermonuclear reactions. 
Because of its fragility, the production of Li in stars remained elusive for decades.
Only in the past ten years or so novae emerged as a serious candidate for the stellar source, supported by observations but not yet by theory. In view of the uncertainties in the production and depletion mechanisms of Li, the evolution of its abundance in a galactic context remains poorly understood, much more so than for any other element.
Those issues - and several others - are the topic of this review. 

\section{LITHIUM AS A PROBE OF STELLAR PHYSICS} 
\label{Observations}

\subsection{The historical constraint of lithium depletion on stellar modelling} 
\label{LiDepletionHistory}

\begin{figure}[h]
\begin{minipage}{.38\textwidth}
{\includegraphics[width=\textwidth]{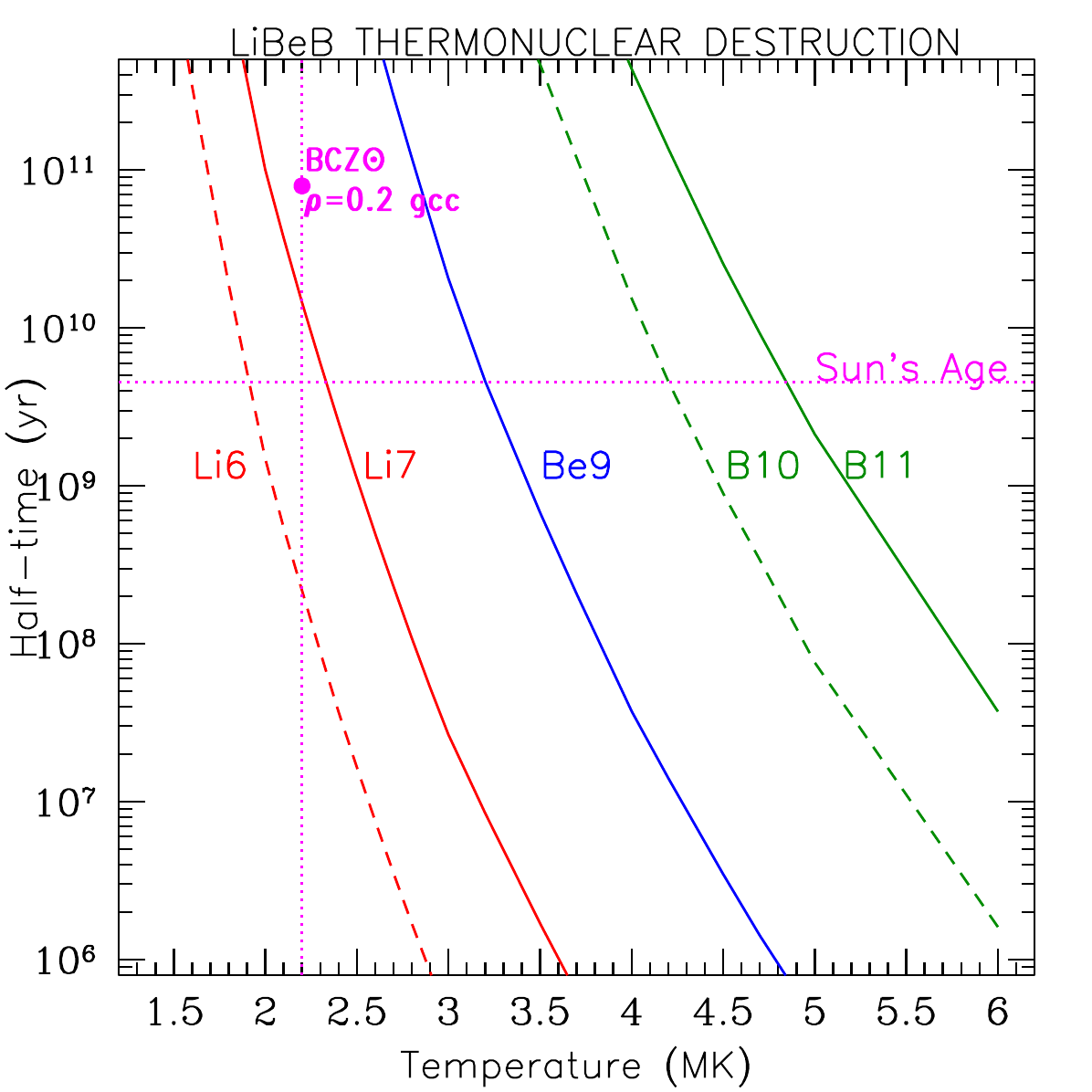}}
\end{minipage}
\hfill 
\begin{minipage}{.66\textwidth}
{\includegraphics[width=\textwidth]
{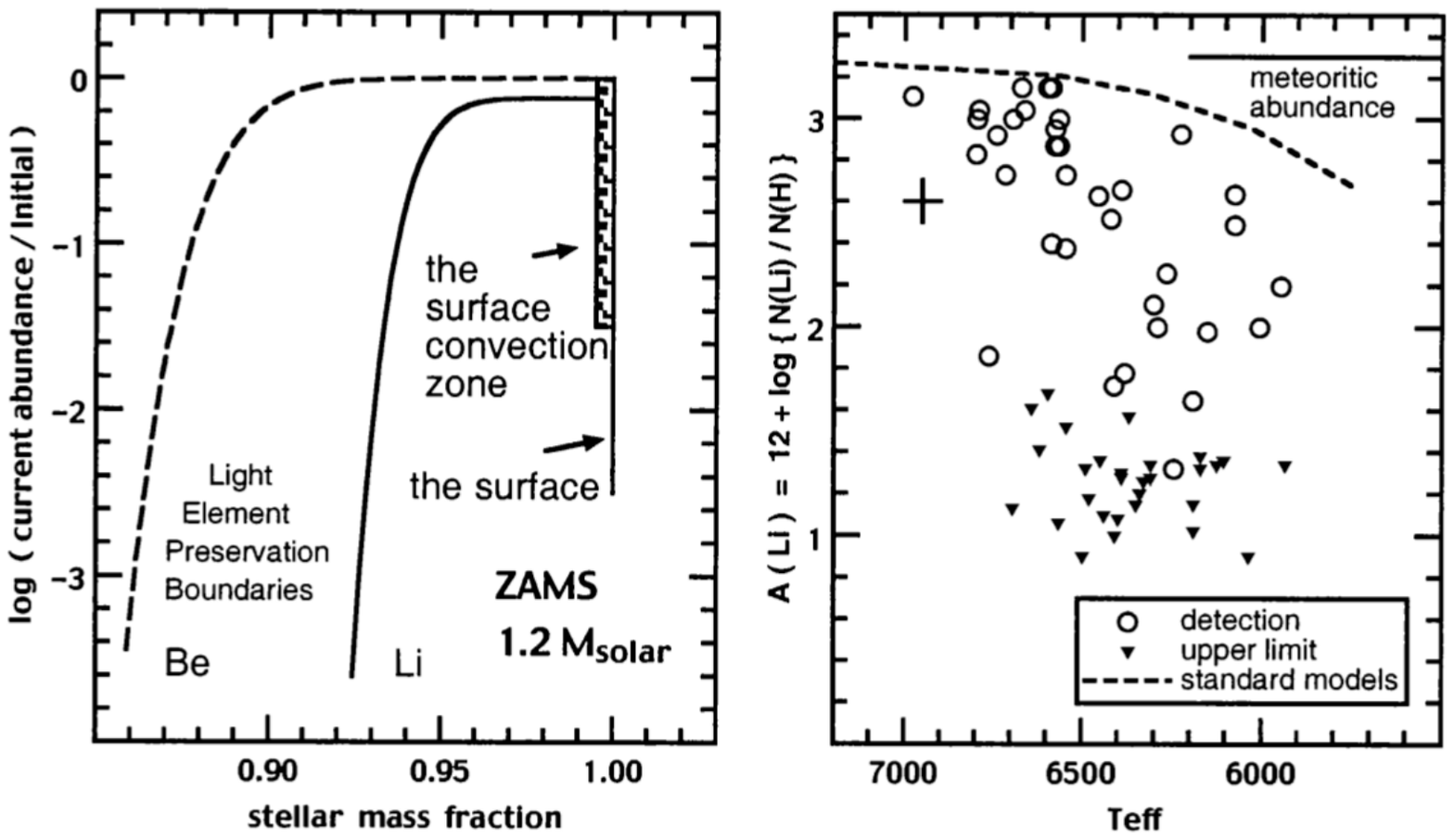}}
\end{minipage}
\caption{ 
{\it Left :} 
Half-times (time of survival of half the original amount) of 
the light isotopes \lia, \lib, \beb, \ba \ and \bb \ as a function of temperature  in an environment of solar composition and density $\rho$=1 \gcc. Timescales for 2-body reactions are inversely proportional to $\rho$. The bottom of the convective zone of the Sun (BCZ$\odot$) has a density of $\rho$=0.2 \gcc, implying a Li burning timescale larger than a Hubble time. 
{\it Middle:} Li and Be preservation zones in a classical stellar model of a 1.2 \ms . The ``step-like" abundance profile results from the temperature gradient in the external layers and the sensitivity to temperature of the proton captures on $^7$Li and $^9$Be \citep[Fig. from][]{2000IAUS..198...61D}. {\it Right:} Predictions of the classical theory, with only pre-main sequence Li depletion in cooler stars \citep[Fig. from][]{2000IAUS..198...61D}.
}
\label{fig_Nucleardestruction-abundanceprofiles}
\end{figure}

In his seminal paper, which established the energy sources of normal (main sequence or MS) stars \cite{Bethe_1939} evaluated the reaction rates of all light nuclei with protons in conditions pertaining to the solar interior. He concluded that, in view of their fragility ``...{\it all the nuclei between H and C, notably H2, H3, Li6, Li7, Be9, B10, B11, can exist in the interior of
stars only to the extent to which they are continuously reformed by nuclear reactions". }
%
%
Indeed, those nuclei are {\it{destroyed}} through (p,$\gamma$) or (p,$\alpha$) reactions in hot stellar interiors on timescales shorter than the stellar lifetimes, as illustrated in  
Fig.~\ref{fig_Nucleardestruction-abundanceprofiles} (left). 
It was soon realised that the strong dependence of the destruction rates of those nuclei on temperature, in conjunction with their abundances, as observed in stellar photospheres,    provides a powerful probe of the inner layers of stars.

\begin{marginnote}
\entry{Li destruction}{Nuclear burning of Li through nuclear reactions with protons}
\entry{Li depletion}{Decrease of Li in the stellar photosphere compare to the original abundance}
\entry{Li dilution}{Decrease of the Li abundance due to dilution with Li-free material during the  dredge-up episodes}
\end{marginnote}

\cite{Greenstein_1951} noticed that Li is depleted in the solar photosphere by a factor of $\sim$100 with respect to the Li meteoritic abundance and attributed that to its destruction through reactions with protons at temperatures T$\sim$3 10$^6$ K (3 MK). They suggested that Li could be brought from the (much cooler) photosphere down to regions of such temperatures either by rotational or convective mixing.  
Subsequently,  \cite{Greenstein_1954} found that Be appears undepleted in the Sun. They suggested that combining these observations one may constrain the depth of the solar convective envelope, since Be destruction by proton capture happens at higher temperature (Fig.~\ref{fig_Nucleardestruction-abundanceprofiles}).  The idea was quantitatively explored by \cite{Salpeter_1955} who used a purely radiative and chemically homogeneous solar model 
and determined that the bottom of the solar convective zone would lie at about half the solar radius, at temperatures between 2.6 and 3 MK. 
However, in the first solar models by \cite{1957ApJ...125..233S} accounting for convective equilibrium in the outer solar layers and chemical inhomogeneity caused by hydrogen burning in its interior, the base of the convective envelope  reached a temperature  of the order of 1~MK only. 
It was then proposed that the solar  lithium depletion might have happened during its pre-main sequence contraction, when the base of its convective envelope was deeper and hotter. This was empirically  proven insufficient by \cite{1963AJ.....68T.280H} and \cite{1965ApJ...141..610W}'s discovery of   G-type main sequence stars with Li abundances similar to the meteoritic values in the young Pleiades OC, and of their slightly more Li-poor counterparts in the slightly older Hyades. This was the first strong evidence for some mechanism(s)   gently transporting Li from the bottom of the convective envelope of solar-type main sequence stars to the deeper radiative layers where it burns. Lithium was then definitively recognised as one of the key 
probes of the interior of low-mass stars, well before the emergence of helio- and asteroseismic constraints. Over the following sixty years, a plethora of observational surprises emerged (Sect.\ref{Depletion-obs} and \ref{Li-obs-evolved}), along with associated theoretical challenges to go beyond classical  stellar modelling (Sect.\ref{LiDepletionTheory}). 
\begin{marginnote}
\entry{Classical stellar model}{Stellar evolution model with convection as the only mechanism  for the mixing of chemicals in a star 
}
\end{marginnote}

\subsection{Observational evidence of Li depletion in low-mass dwarf stars} 
\label{Depletion-obs}

\begin{marginnote}
\entry{A(Li)} {$\rm {log10(N_{Li}/N_H) + 12}$ where $N_{\rm Li}$ and N$\rm{_H}$ are the number density
of lithium and hydrogen respectively}
\end{marginnote}

\subsubsection{The Sun and solar analogues}
\label{Depletion-obs-solartype}
 
A significant research effort has been dedicated to the reconstruction of the history of lithium depletion in the Sun, with a particular emphasis on solar analogues. 
These dwarf stars are similar to the Sun in terms of mass ($1 \pm 0.1$~M$_{\odot}$) and metallicity ([Fe/H]$=0 \pm 0.2$). However, they are possibly of different ages and belong either to OCs or to the field.
It is now well established that the photospheric Li abundance of these objects decreases along the main sequence, 
with indications that slower rotators and/or more metallic solar analogues exhibit lower Li content \citep{1993AJ....106.1059S,1997AJ....113.1871K,2007A&A...468..663T,2013ApJ...771L..31D,
2013ApJ...774L..32M,2014A&A...567L...3M,
2023MNRAS.522.3217M,2024AJ....168..240L}. Important Li scatter among field solar analogues at a given age was usually reported, with some suspicion  that part of it was due to small differences in mass, metallicity, and evolution status of the sample stars, impacting their Li-depletion history. In particular, based only on spectroscopy and on the isochronal method with very limited applicability to dwarfs \citep[][and references therein]{2014EAS....65..177L,2021A&A...646A..78M}, 
the Sun has long been thought to have the lowest Li abundance among field solar twins of similar age \citep[][and references therein]
{
2020MNRAS.492..245C}.  
This was demonstrated erroneous, thanks to asteroseismology which provides precise and accurate values for the age, mass and radius of oscillating stars \citep{2014ApJS..210....1C,2021RvMP...93a5001A,2021A&A...646A..78M,
2021MNRAS.505.2151C}. The combined spectroscopic, photometric, and seismic analysis of solar analogues observed with {\it{Kepler}}, which also provided their surface rotation periods (hence alleviating the observational restriction related to the spectroscopic determination of vsin$i$), demonstrated that the solar Li and rotation rate appear normal for a star like the Sun at its age \citep{2017A&A...602A..63B}.

\begin{marginnote}
\entry{Population I}{Pop I stars are typically younger than 10 Gyr, metal rich (\feh$>$-1), and located near the plane of the Galactic disc
}
\entry{Population II}{Pop~II stars are 11-13.5 Gyr old, metal poor (\feh$<$-1), and are found in the Galactic halo}
\end{marginnote}

\subsubsection{Dwarf stars in open clusters}
\label{Depletion-obs-OC}

\begin{figure}[t]
{\includegraphics[width=1.\textwidth]{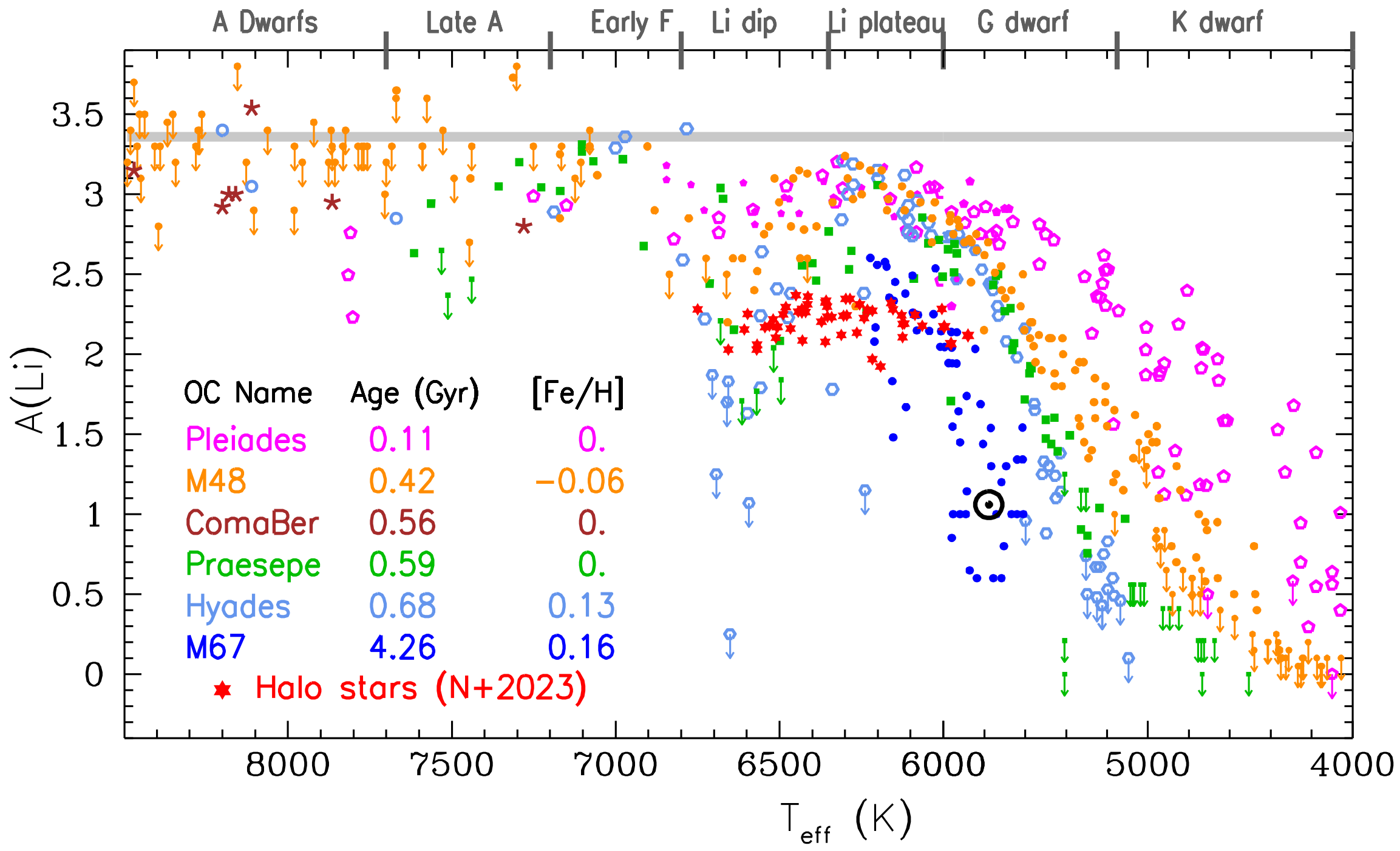}}
\caption{Li abundances in OC and halo dwarf stars as a function of effective temperature. The stellar types and the main Li abundance patterns discussed in the text are indicated. 
The OC data are: for Pleiades from \citet[][filled magenta]{Boesgaard2003a}  and \citet[][open magenta]{Bouvier2018}, for M67 from \citet[][filled blue]{Pace2012}, for Praesepe from \citet[][green squares]{2017AJ....153..128C}, for Coma Ber from \citet[][brown asterisks]{2000A&A...354..216B} and for M48 from \citet[][orange]{2023ApJ...952...71S}.
The Li abundances of  Pop~II halo 
dwarfs are from the \citet{Norris_2023} sample, as cleaned in \citet{Borisov2024}. The grey line indicates the meteoritic Li abundance. The temperature scales vary among the studies and 
have not been homogeneised  here.} 
\label{fig_Li-teff-morphology}
\end{figure}

Open cluster (OC)  
stars are supposed to be coeval and to share the same initial composition, with both age and composition varying from one cluster to another. Starting with the seminal work of \citet{1965ApJ...141..588H} and \citet{1965ApJ...141..610W},
their study has depicted the Li depletion history in Pop~I dwarf stars, and demonstrated its dependence with spectral type.
The well-known and ubiquitous Li-\teff-age patterns for AFGK dwarf stars are clearly identifiable in Fig.~\ref{fig_Li-teff-morphology} that shows Li data in several OC of different ages and [Fe/H]. The effective temperature is a proxy for stellar mass at a given metallicity and age, 
starting from hotter and more massive A- and  F-type stars on the left of the plot and moving to cooler and less massive G- and K-type stars towards the right. 

The highest Li abundances 
 are found in the youngest systems, as well as in A- and F-type stars with T$_{\rm eff} \geq 6800 K$ \citep[][]{1986ApJ...302L..49B,1986ApJ...309L..17H,1989A&A...220..197B,
 2000A&A...354..216B,1997A&A...323...86R,Randich2020,
 2023ApJ...952...71S}. Those 
 lie along a plateau close to the  
 meteoritic value, independently of the age of the OC. 
 Relatively few abundance
determinations are available in this \teff \ range, as 
the primary abundance diagnostic, the Li I 670.8 nm resonance
line, is weaker in hotter stars. 
Excessive line broadening further complicates spectroscopic
analyses in fast rotators ($v$sin$i$ $> 25$km s$^{-1}$),  which constitute the majority of A- and F-type stars. This explains why in Fig.\ref{fig_Li-teff-morphology} most of the A-type stars in M48 from \citet{Sun_2025}'s sample have Li upper limits.  
 Cooler dwarf  stars exhibit lower levels of lithium in their photospheres as OC age.

 A drop-off in the Li content centred around 6500-6700~K is clearly visible in Fig.~\ref{fig_Li-teff-morphology}. The so-called Li dip 
 appears in  OC older than $\sim$ 150-200 Myrs \citep{1965ApJ...141..610W,1986ApJ...302L..49B,1988ApJ...334..734H,
 1995ApJ...446..203B,2000A&A...354..216B,2001A&A...374.1017P,
 2004ApJ...614L..65S} but not in younger ones 
 \citep[][and references therein]{2023ApJ...952...71S}.
 No significant alteration of the Li distribution within the dip is detectable in clusters older than the Hyades \citep[$\sim 700$~Myrs; ][]{2009AJ....138.1171A}. This pattern thus builds up early on the main sequence. Importantly, the dip's position occurs at nearly identical \teff ~across a wide range of metallicity, and corresponds to higher mass stars in higher metallicity OC \citep{2012AJ....144..137C,2013A&A...552A.136F}.
 In older OC like M67 or NGC~3680, Li dip stars have left the main sequence, but the analysis of slightly evolved subgiants originating from the \teff ~location of the dip confirms that this pattern is universal \citep{1995ApJ...446..203B,
 2001A&A...374.1017P,2019AJ....158..163D}. 
 
\begin{marginnote}
\entry{Li dip}{Drop-off in the Li photospheric abundances of F-type main sequence stars centred around 6500-6700~K and 
observed at all [Fe/H]$\geq -1$}
\end{marginnote}
 
On the cool side of the dip, Li rises with a tight Li-\teff \ relation  \citep[with only a few binaries falling below,][]{2017AJ....153..128C} up to another plateau (between $\sim 6350$ and $\sim 6000 K$), hereafter the Pop~I Li plateau.  
 Its Li value decreases uniformly with age and is lower than the plateau above the high-temperature boundary of the Li-dip.  

 Moving towards lower effective temperature, 
 lower mass 
 late-G and K-type stars  
 exhibit Li depletion already in the youngest OC before their arrival on the zero age main sequence (ZAMS; i.e., those stars undergo moderate Li depletion on the pre-MS, see the data for NGC~2547 and the Pleiades in Fig.~\ref{fig_Li-teff-morphology}), as well as later on the main sequence. The Li depletion levels are strongly \teff-dependent at all ages, with steeper Li-\teff-depletion trend in more metal-rich clusters at a given age  \citep{1972ApJ...172...57Z,1984ApJ...283..205C,2005A&A...442..615S,2012AJ....144..137C}. Finally, and in contrast to the findings observed on the cool side of the Li-dip, G-dwarf binaries align perfectly with the trends observed for single stars (not shown in Fig.~\ref{fig_Li-teff-morphology}, but see \citealt{2023ApJ...952...71S}).
 \begin{marginnote}
\entry{Pop~I Li plateau}{Relatively constant Li abundance for Pop~I dwarf stars with \teff between $\sim$ 6000 and 6350~K.}
\entry{Pop~II Li plateau}{Relatively constant Li abundance in the Galactic halo and globular cluster subdwarf stars (including extra-galactic accreted ones) with  
6000 $\leq$ T$_{\rm eff}/{\rm K} \leq$ 6350 and [Fe/H] $\leq$-0.7}
\end{marginnote} 

\subsubsection{Population~I field dwarf stars} 
\label{Depletion-obs-PopIFieldDwarfs}
The Li-\teff-age patterns observed in OC have been clearly identified in Pop~I field stars, first from small stellar samples  \citep{1990ApJ...354..310B,1991MNRAS.253..610L} and more recently with large spectroscopic surveys.
The Li dip was clearly identified from GALAH data 
as a significant overdensity of 
stars with Li upper limits in the specific region of the Kiel diagram spanning the evolution path of Li-dip stars (\teff \ between $\sim$ 6300 and 6600~K, surface gravity logg between 3.8 and 4.3, 
[Fe/H] between -1 and +0.5; \citealt{Gao2020}). 
It is thus clearly established that the Li dip is a universal main sequence phenomenon and that it is confined to the same effective temperature range for all metallicities at which it is potentially observable (i.e., down to [Fe/H]$\sim -1$; more metal-poor stars originating from the hot side of the dip have already left the main sequence). In addition, \citet{Gao2020}  retrieved 
the different Li-depletion regimes for the groups of main sequence stars that are, respectively, cooler and hotter than the Li dip. The warm group exhibits no significant Li depletion as they age along the main sequence, independently of their metallicity, while the Li depletion in the cool group is strongly dependent on age, \teff, and metallicity. 

\subsubsection{Observational evidence of Li depletion in Population~II Galactic and extra-galactic dwarf stars} 
\label{Depletion-obs-PopIIhalodwarfs}

The discovery by \citet{Spite_1982} of a relatively constant Li abundance  
in Pop~II halo subdwarf stars with [Fe/H] between -0.7 and -2.5 and \teff \ between $\sim 5800$~K (or 6000~K, depending on the adopted temperature scale) and 6300~K (6500~K) has opened up a large number of discussions linking BBN (\S~\ref{subsec_bigbang}), Galactic chemical evolution (\S~\ref{subsec:Proto_solarLi_and_mainsource}), and stellar physics (\S~\ref{LiDepletionTheory}). 
Several surveys of Li in MS turnoff halo stars 
have investigated what is known as the Spite Li plateau (Fig.~\ref{fig_Li-teff-morphology} and \ref{fig_Li_depletion}), with follow up stretching down to [Fe/H]$\sim$-6
\citep{1991ApJ...375..116H,1994ApJ...421..318T,
1994ApJ...423..386N,Norris_2023,1995A&A...295L..47M,
1996ApJ...458..543R,Ryan1999,2001ApJ...549...55R,
1996A&A...307..172S,1997MNRAS.285..847B,Charbonnel2005,2008ApJ...684..588F,2019ApJ...871..146F,
Sbordone2010,2010A&A...511A..47H,2011Natur.477...67C,2012A&A...542A..87B,2015A&A...579A..28B,2017AJ....154...52M,
2019ApJ...874L..21A} with an overview of the results presented in \citet{Bonifacio_2025}. 
Their findings have prompted extensive debate about the actual  abundance level of the Pop~II plateau, with a general agreement that it lies roughly a factor of 3 below the theoretical BBN primordial Li abundance (\S~\ref{subsec_bigbang})\footnote{For dwarf stars with \teff$>$6000~K and -2$<$[Fe/H]$<$-1, \citet{Norris_2023} derive $<$A(Li)$>$ = 2.322$\pm$ 0.013, with $\sigma$(A(Li)) = 0.078.}, its thickness, the dependencies of Li on \teff \ and [Fe/H], the apparent meltdown at [Fe/H] below -3 and the evidence for an  anticorellation between Li and [C/Fe] in the most extremely metal-poor stars. 
Prior to the review of these points, it is imperative to emphasise that the accurate determination of the actual characteristics of the Li plateau is fundamental for several reasons. 
On one hand, any possible Li trend with metallicity is important for evaluating the Galactic chemical evolution to the observed Li abundances in stars (\S~\ref{subsec:Proto_solarLi_and_mainsource}), and it can also contribute to a scatter in the data at fixed effective temperature. On the other hand, the level of the plateau as well as any dispersion and/or trend with effective temperature at different metallicities provide key  diagnostics on the physical processes that can lead to Li depletion in halo stars 
with respect to the initial, potentially primordial Li they were born with (\S~\ref{subsec_bigbang}). 
We refer to \citet{Charbonnel2005} and \citet[][and references therein]{Norris_2023} 
for discussions about the origin of the observational discrepancies found in the literature and their relations to the assumptions made in the different studies for the  determination of the stellar parameters and abundances (e.g. adopted temperature scale, reddening, NLTE and 1-3D corrections). 
 The additional fundamental 
 issue we want to emphasize here is that the derivation of both the Li dispersion and the trends with [Fe/H] and \teff \ (treated separatly or in a bivariate analysis) is sensitive to the selection of the stellar samples and to the treatment of the Li ``outliers" when drawing conclusions.    
This is where 
astrometry and additional spectroscopic information become crucial 
in order to address questions relating to the Li plateau, as illustrated by a couple of examples below.

Using Hipparcos parallaxes to determine the evolution stage of a halo sample from the literature, \citet{Charbonnel2005} showed that most stars with Li under-abundances 
below the plateau in the [Fe/H] range between -1 and -3.5 
were actually slightly evolved, post turnoff stars originating from the hot side of the plateau, whose inclusion resulted in an artificial increase in the Li dispersion previously derived for turnoff stars. 
More recently, using Gaia DR3 to characterise halo dwarf stars from GALAH DR3 \citep{Buder2021} combined with the literature sample homogeneously analysed by \citet{Norris_2023},  
\cite{Borisov2024} confirmed that the Li plateau extends down to [Fe/H]=-6 for stars with \teff ~above 5800~K, with all the Li deficient stars being peculiar anomalous objects. 
\begin{marginnote}
\entry {CEMP stars} 
{Carbon-enhanced metal-poor stars, a class of chemically peculiar stars with \feh$<$-1 and [C/Fe]$>$+1}
\end{marginnote}
In particular, and as underlined by \citet{Norris_2023}, the most metal-poor ([Fe/H] $\leq$-4.5) stars of the currently available samples are both Li-poor (A(Li) between 0.5 and 2.2) and extremely carbon-rich (CEMP with [C/Fe]$>3.5$), 
with an anticorrelation between Li and C indicating that they were probably already Li-deficient at birth due to close environmental effects. 
The apparent Li meltdown in the [Fe/H] range between -4.2 and -3 is also accompanied by C enrichment, with [C/Fe] up to +3.2. These stars are thus not relevant to the original Li abundance patterns at the earliest times in the young Milky Way. When CEMP-s and CEMP-no stars are excluded from the analysis, there is no (or only very little) evidence for a bending of the Li plateau at low metallicity. 
Instead, the constancy of the plateau in the most metal-poor regime is sustained by the finding of the most Fe-poor ($<$ -5.8) Li-plateau star, J0023+0307, which is not a CEMP star \citep{2019ApJ...874L..21A,2019ApJ...871..146F}. Although they should be very rare, we are anticipating that the Li abundance in extremely metal-poor dwarfs with \teff \ above $\sim$6000~K and no significant C-enrichment should lie on the Li plateau. Finally, studies of turnoff globular cluster stars reported Li abundance in these objects consistent with the halo plateau value \citep{2009A&A...503..545L,2024MNRAS.52712120N}. Any successful theoretical explanation of Li depletion must explain both halo and GC multiple stellar populations\footnote{Due to lack of space we do not tackle the question of Li in the multiple stellar populations of globular clusters and we refer to \citet{2025MNRAS.tmp.1257G} for references and discussion.}. 

The observation of Li in unevolved stars in galaxies outside the Milky Way is, at present, unfeasible. 
However, successive mergers with dwarf galaxies have brought some of these originally  extragalactic stars into the Galactic halo, giving access to extragalactic Li abundances. 
The first extra-galactic Li plateau was evidenced 
in the globular cluster-like stellar system Omega Cen, which is thought to be the stripped  core of a dwarf galaxy. The Omega Cen plateau lies at the same Li abundance level than in {\it{in situ}} Galactic halo stars, and it spans a wide range of metallicities that overlap with that of the Spite plateau \citep{2010A&A...519L...3M}. 
Thanks to the extraordinary performance of Gaia astrometry and associated spectroscopic and deep photometric surveys, the same Li plateau was found in  {\it{Gaia}}-Sausage-Enceladus, a  disrupted dwarf galaxy after collision with the Milky
Way about 10 Gyr ago \citep{Molaro2020,2021MNRAS.507...43S}, 
and in the S2-Stream \citep{2021MNRAS.500..889A}. The universality of the Li plateau for [Fe/H]  lower than $\sim -1.0$ 
in other galaxies, regardless of their type and their likely  different star formation and chemical enrichment histories, implies that this feature  does not result from  environmental effects. 
This excludes scenarios  calling for  the astration 
of a large fraction of the interstellar medium (ISM) of galactic haloes by a first generation of massive stars that would have effectively destroyed Li before the formation of Galactic and extragalactic halo dwarf stars  observed today \citep{Piau2006,2025arXiv250809821M}. 
In addition, 
such a process would lead to an important early overproduction of CNO elements, which is not observed in halo stars \citep{PrantzosN_2010} and to a strong depletion of the much more fragile deuterium in the Galaxy, which is not observed either.
\begin{marginnote}
\entry{Astration}{The modification of the composition of interstellar gas after it is re-ejected by a star }   \end{marginnote} 

\subsection{Observational evidence of Li dilution and depletion in evolved low-mass stars, and the case of Li-rich red giants}
\label{Li-obs-evolved}

\subsubsection{Li dilution and depletion in evolved low-mass stars}
\label{Li-depletion-obs-evolved}

When low-mass stars leave the main sequence, they evolve towards the subgiant and red giant phases. Their radius increases, they become redder and brighter, and their surface rotation drops, facilitating the Li abundance determination in these objects 
\citep{1940PASP...52..407M}.
On the other hand, their convective envelope deepens towards internal layers where Li was fully destroyed by proton captures in the earlier phases of their evolution. Consequently, the surface Li abundance decreases due to dilution with Li-free material during the so-called first dredge-up (1DUP; \citealt{1967ApJ...147..624I}; Sect.\ref{theoryadvancedphases}; \citealt{1940PASP...52..407M,1975ApJ...195..649A,1980ApJ...235..114L,1999A&A...345..936L,
2003A&A...409..251M,
2011A&A...527A..94C}
). During that phase, the convective envelope also  dredges-up deep layers where the CN-branch 
occurred during the main sequence, leading to a decrease of the surface C/N and carbon isotopic ratios \citep{1994A&A...282..811C}.  
These LiCN evolutionary dilution  patterns were clearly identified among FGK field red giants over a wide metallicity range \citep{1993PASP..105..337S,2023A&A...670A..73A
}, as well as in open clusters \citep{1986ApJ...300..289P,
1986ApJ...301..860S,1988PASP..100..474P,
1989ApJ...347..835G,2004A&A...424..951P,
2015MNRAS.446.3562B,2016A&A...591A..62S,2018AJ....155..138A,
2019AJ....158..163D
} and in globular clusters  \citep{2009A&A...503..545L,2021A&A...653A..13A}. 

After the end of the 1DUP, the surface Li abundance stays constant as the low-mass stars climb the low luminosity part of the RGB, until they reach the location of the so-called RGB bump. There, a second drop of the Li surface abundance occurs, simultaneous to a second drop of the carbon isotopic ratio, when the corresponding data are available 
\citep[][for both Pop~I and Pop~~II (including GC) giants]{1998A&A...332..204C,2000A&A...354..169G,2009A&A...503..545L}.
Both the Li depletion and the decrease of the carbon isotopic ratios are observed to be more drastic when the mass and/or the metallicity of the RGB stars are lower. 
This trend is corroborated by asteroseismology coupled to spectroscopy \citep{2024MNRAS.527.8535C,2024A&A...684A..70L}. 
More insight on the evolution of Li in evolved stars of different initial masses comes from large spectroscopic surveys combined with Gaia data, which confirm the global trends described above  \citep{2020A&A...633A..34C,2021A&A...651A..84M,Wang_2024}. This includes the discovery of Li-rich giants discussed below, the nature 
of which  
is strongly  debated.

\subsubsection{Li-rich red giants}
\label{Li-rich-red-giants-obs}
In principle, the criteria for classifying a giant star as Li-rich is very simple. It is a star that is evolved enough to have  completed the 1DUP, 
and which exhibits a photospheric Li abundance significantly higher than  
that of its evolved counterparts as well as theoretical predictions at the same evolution stage.  Historically, Li-rich TP-AGB stars have been treated separately (Sect.\ref{subsec:HBB_AGB}), and the stars we discuss here are potentially on the RGB, in the red clump or the horizontal branch, or on the early-AGB. 
Following the unexpected discovery of some of these objects \citep{1982ApJ...255..577W}, systematic surveys were carried out to search for them among field giants \citep{1989ApJS...71..293B,1999A&A...352..495M,
1999A&A...342..831J,2011ApJ...730L..12K,2013MNRAS.430..611M},   
with a recent acceleration in the domain thanks to 
Gaia-ESO, GALAH and LAMOST  \citep{2016MNRAS.461.3336C,Gao_2019,2023AJ....165...52C,
2024ApJS..271...58D}. 
In some cases,  the selection function or the observing strategy have  precluded the precise determination of the frequency of Li-rich red giants and its possible dependence with  parameters like the stellar mass or metallicity. However, what is usually found in the literature is that $\sim 1 - 2 \%$
of field G and K giants have enhanced photospheric lithium
abundances with respect to their Li-depleted counterparts. Of those, 
$\sim 3\%$ 
have Li above the meteoritic value, with only a
handful exhibiting A(Li)$>4$. 

These estimates and their interpretation about the nature of the (super-) Li-rich giants should be taken with great caution. 
The popular Li threshold value of 1.5  corresponds roughly to 
 the canonical post-1DUP value predicted by classical models of solar metallicity stars with masses $\sim 1.25-1.5$M$_{\odot}$ \citep{1967ApJ...147..624I}, assuming they all formed and left the main sequence with the Li meteoritic value. This  simplification has resulted in a significant amount of confusion in the identification of giants that are really abnormally rich in Li, for the following  reasons. First and as explained in  Sect.~\ref{Depletion-obs}, stars don't leave the main sequence with their original Li photospheric abundance, and Li depletion on the main sequence is both mass and metallicity dependent; therefore, post-1DUP Li abundances are expected to vary accordingly. 
 Second, isochronal determination of the mass  of red giants  applied in most studies is highly uncertain and the asteroseismic mass-determination is still  limited to small number of stars  \citep{
 2021A&ARv..29....4S,2024AJ....167..208W,2025MNRAS.541.2631C}.   Nevertheless, asteroseismology facilitates the differentiation of red clump stars undergoing central He-burning from  RGB and early-AGB stars that are undergoing H- (and eventually He-) shell burning \citep{2011Natur.471..608B,2025A&A...697A.165V}.
 Last but not least, the distribution of the masses of today's giants in any field star  sample is different in the different regions of the HRD, and it also depends on the metallicity and the age of the populations investigated.   
 Consequently, the interpretation of Li data from large surveys requires the understanding of the mass distribution and a careful consideration of the selection effects. 
This is elegantly illustrated by 
\citet{2022ApJ...933...58C} who  revisited the Li data  
from GALAH DR2 for evolved stars.  
Assuming that all red clump (RC) stars from the same sample have masses around 1~M$_{\odot}$,  \citet{2020NatAs...4.1059K}  had previously claimed that all these stars have high levels of Li for their evolution stage (i.e., are Li-rich), calling for a Li production phase between the RGB tip and the clump. 
\citet{2022ApJ...933...58C}  demonstrated, however, that 
the Li abundances observed in RC stars are  normal when one considers a realistic stellar mass distribution in the sample together with a proper mass-dependent Li depletion on the main sequence and the corresponding post-main sequence dilution. 
More precisely, they predicted  
a large fraction of very low
Li abundances (hence a large number of upper limits) from low-mass RC progenitors, and higher Li abundances from higher mass ones. These trends have been confirmed with GALAH DR4 \citep{Buder2025}. 
Therefore, their conclusion  that {\it{``there is no evidence among clump giants of an unknown Li production mechanism occurring between the upper RGB and the horizontal branch for low-mass stars"}}, at odds with \citet{2020NatAs...4.1059K}'s conclusion, is robust. 

Given the issues and the example discussed above, important efforts must be devoted to ascertain the mass and the position in the HRD diagram of  Li-rich and super Li-rich giant candidates,  potentially with asteroseismic clues on  their evolution phase, and with the determination of their carbon isotopic ratio, which is an additional  indicator of these factors \citep[][and references therein]{2024A&A...684A..70L}. The improvement in the determination of the parallaxes and luminosities thanks to Hipparcos and Gaia astrometry provided some evidence that Li-rich and
super Li-rich cool and low-mass giants (typically below $\sim$ 2~M$_{\odot}$) are primarily located in specific regions of the colour-magnitude diagram, namely close to the RGB bump or on the early-AGB \citep{2000A&A...359..563C,2011ApJ...730L..12K,
2024ApJ...973..125K}. 
When available, the low values of $^{12}$C/$^{13}$C found in these objects  corroborate this finding \citep{
2000ApJ...542..978B,2018A&A...615A..74Z}, suggesting  internal Li production to replenish Li at the surface of red giants during very brief evolution phases.
In a few OC, Li-rich giants with similar CMD location have been found 
\citep{2013ApJ...767L..19A}. 
Conversely, some studies have identified Li-rich giants in various locations along the RGB, which lends support to the hypothesis of an external contribution of Li, possibly resulting from the engulfment of sub-stellar objects 
\citep{
2016MNRAS.461.3336C,2016A&A...587A..66D,2016ApJ...833L..24A}. 
 This phenomenology should be scrutinized further through a fully consistent analysis of  the available data accounting for all the possible observational 
 and sampling biases, including cluster membership. 

We conclude with 2MASS J05241392-0336543, which is the most Li-enhanced giant star discovered to date (A(Li), 3D, NLTE = 6.15 $\pm$ 0.25; 
[Li/Fe]=+7.64 $\pm$ 0.25; \citealt{2024ApJ...973..125K}). 
It is  very metal-poor ([Fe/H] = -2.43$\pm$0.16), C-enhanced ([C/Fe]$=+0.51$), and it has a low $^{12}$C/$^{13}$C ($10 \pm3$). 
Given its relatively low \teff, 
it is definitively not undergoing central He-burning on the horizontal branch (the equivalent of the clump for metal-poor
giants).  
Instead, it is either a RGB star located just above the bump, or on the early
asymptotic giant branch (e-AGB). This provides evidence for fresh $^7$Li production and excludes both preservation of primordial Li and planetary accretion as viable scenarios for the formation of this red giant star with the highest ultralithium enhancement.  

\section{HOW DO STARS DEPLETE LITHIUM}
\label{LiDepletionTheory}

{\it{``The history of our attempts to understand the abundance of lithium in the Sun has had its share of vicissitudes" \citep{1965ApJ...142..174W}}}. 
After the acknowledgement, in 1965, 
that the Li depletion history in the Sun and in low-mass dwarf stars could not be explained by the so-called classical theory of stellar evolution  
(in brief, the surface convection zone is much shallower than the Li preservation layers; 
Sect.~\ref{LiDepletionHistory} and Fig.~\ref{fig_Li-teff-morphology}), 
several physical processes have been considered 
for reproducing the observations. 
The theoretical challenge 
was (and still is) threefold: 1) to identify 
the physical processes that transport chemicals and angular momentum in stellar interiors, and their possible interactions; 2) to describe them in a way that could be portable into 1D stellar evolution codes; and 3) to test them consistently across the HRD, both in terms of spectral type  and evolution phases, and over the metallicity ranges where data are available. We focus on those mechanisms whose consideration to solve self-consistently the stellar Li puzzles appears to 
still hold relevance today. 

\subsection{Atomic diffusion and competing processes} 
\label{subsec:atomicdiffusion}

Atomic diffusion of different species against the hydrogen background 
is mainly led by the gradients of pressure or gravity, of concentration, and of temperature, and by radiative acceleration, 
which can 
be computed from first principles \citep[][and references therein]{
2015ads..book.....M}.
It is considered as a ``standard" process 
in the modelling of low-mass stars, 
as it lowers their lifetime \citep{2002ApJ...571..487V}. 
It is however widely recognized that, in the vast majority of stars, including the Sun as supported by helioseismology \citep[][and references therein]{2021LRSP...18....2C},  
the effects of atomic diffusion are smoothed out, at least partly, by  competing macroscopic processes. Li has played (and is still playing) a key role in their identification.

\begin{figure}[t]
\begin{center}
{\includegraphics[width=0.7\textwidth]{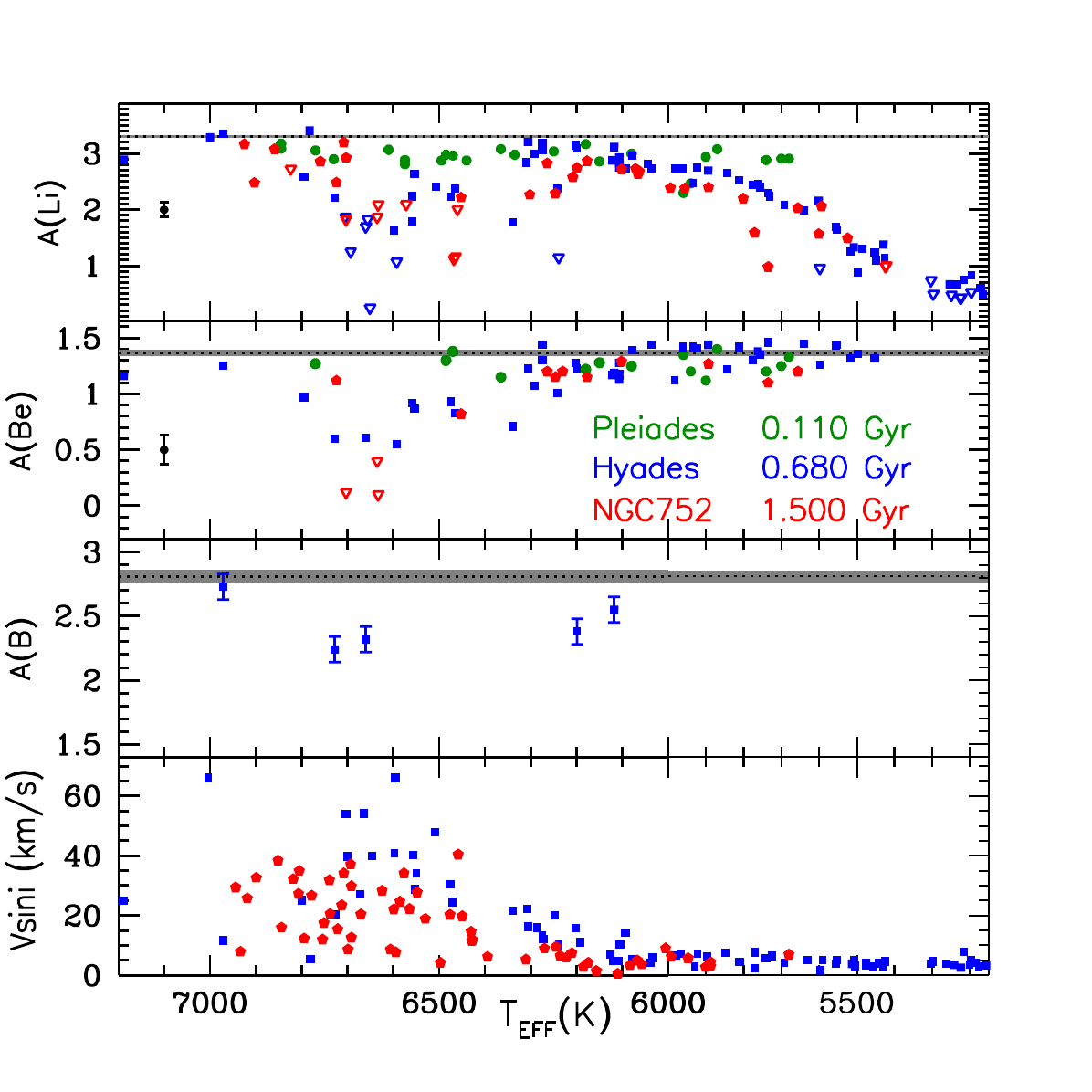}}
\caption{Abundances of Li ({\it top}), Be ({\it second from top}), B ({\it third from top}) and $\upsilon sini$ ({\it bottom}) versus T$_{\rm eff}$ for dwarf stars in the Pleiades (green dots), Hyades (blue squares) and NGC752 (red pentagons), with upper limits indicated by open downturned triangles. Horizontal grey lines indicate proto-solar values. Figure adapted from  \citet{2023ApJ...943...40B} and references therein.}
\end{center}
\label{LiBeBdipVsini}
\end{figure}

If the radiative layers underneath the convective envelope of dwarf stars were totally stable, the efficiency of 
atomic diffusion increases (i.e., its timescale decreases) with decreasing density, hence with increasing stellar radius.  
 Hence, while the timescale for atomic diffusion in the Sun is too long to explain the solar Li, 
it should significantly alter the photospheric composition of hotter stars. In particular, the external H-He convection envelope gets rapidly shallower as \teff  ~increases between $\sim$~6000 and 7000~K. This led to  \citet{1986ApJ...302..650M}'s original suggestion that the Li dip centered around 6700~K could be shaped by atomic diffusion, with gravitational settling causing the drop in Li on the cool side of the dip as the stellar convection envelope shrinks, and radiative levitation or mass loss leading to the rise on the hot side of the dip.
Three serious drawbacks were however quickly identified. First, the expected concomitant abundance variations of heavier elements (C, N, O, Mg, Si, Cr, Mn) 
are not observed across the \teff ~range of the Li dip 
\citep{1993ApJ...412..173G,
1999A&A...351..247V,
2013PASJ...65...53T}.
Second, the diffusion timescales would lead Li to settle and accumulate in a buffer zone below the convection envelope; 
Li should thus be dredged-up when the stars enter the Hertzsprung gap, which is not observed in field and open cluster subgiants \citep{1988PASP..100..474P,2000A&A...357..931D,2025arXiv250704266S}. 
Third, in the case of pure atomic diffusion, Li, Be, and B would sink below the convective envelope at very similar rates. While a Be-dip was evidenced in several OC at the same \teff ~as the Li dip, it is much shallower,  
and B is hardly depleted in Li and Be dip stars \citep[Fig.\ref{LiBeBdipVsini}; ][and references therein]{2022ApJ...927..118B}. 
Explanations relying on a transport process leading to the nuclear destruction of Li and Be in 
F-type stars are thus favoured, with strong constraints on its efficiency and radial extension coming from the Li/Be/B ratio inside the dip. A substantial body of evidence has emerged indicating that rotation-induced processes are likely the culprit. 

Another sign that some mild transport process counteracts atomic diffusion in low-mass dwarf stars of similar effective temperature, whatever their metallicity, is the flatness of the Pop~II Li plateau (Sect.~\ref{Depletion-obs-PopIIhalodwarfs}). With atomic diffusion alone, the Li depletion would increase with effective temperature \citep{1988ApJ...335..971V,1990ApJS...73...21D,
1991ApJ...371..584P}. The introduction in stellar evolution models of a very simple turbulent transport coefficient 
in the radiative interior is all that is required to produce a  flat Li abundance plateau as observed in Pop~II dwarfs \citep{2002ApJ...568..979R,Richard2005}. When the parameters describing mild turbulence are adjusted to reproduce the small abundance variations of metals (Mg, Ca, Ti, Fe) between turnoff and subgiant stars in globular clusters, Li depletion by $\sim$ 0.2 to 0.5 from its original abundance is predicted \citep{2006Natur.442..657K,2016A&A...589A..61G,2024MNRAS.52712120N,Borisov2024}. There are numerous indications that rotation is at the origin of the mixing process involved. 

\subsection{Type I rotating models}  
\label{TypeImodels}

Rotation was suggested to play a dominant role in shaping the Li behaviour in low-mass stars since  
\citet{1987PASP...99.1067B} noticed that the \teff ~of the Li dip in the Hyades is  associated with the  drop in rotation velocities that was first evidenced by \citet[][
]{1967ApJ...150..551K}, as  illustrated in Fig.~\ref{LiBeBdipVsini}. This sharp decline in rotational velocities of main-sequence stars later than spectral
type F4V is related to the transition between hotter stars
with envelopes in radiative equilibrium and cooler stars with well-developed subsurface H-He
convection zones.   
In the latter ones, 
magnetically coupled winds are responsible for the deceleration observed early on the sequence \citep[][and more details below]{1966ApJ...144..695W}. 
The coincidence  between the Li and Be dips and the rotation drop (i.e., the efficient extraction of angular momentum by the stellar winds) was confirmed in several OC \citep{2016ApJ...830...49B,
2017AJ....153..128C}.

\begin{marginnote}
\entry{Type I rotating models} {Models of rotating stars where the transport of angular momentum and of chemicals is driven by meridional circulation and shear turbulence.}
\entry{Type II rotating models}{Models with additionally considered magnetic fields and/or internal gravity waves for the transport of angular momentum and chemicals}
\end{marginnote}

Li has  played a key role to test the so-called Type I rotating models that are based on \citet{Zahn1992}'s pioneering  developments 
\citep[][and references therein]{2022csss.confE..92P}.
In this framework, the  modelling of rotation-induced mixing is inextricably linked to the modelling of the angular momentum transport  within the star, which is dominated by the Eddington-Sweet (large scale) meridional circulation and by shear instabilities. The circulation is driven by the loss (or the gain) of angular momentum through  magnetic winds (or through accretion). The resultant internal rotation is  non uniform; a baroclinic state ensues, with the temperature varying with latitude along isobars. Additionally, the differential rotation is expected to trigger hydrodynamical turbulence through various instabilities (shear, baroclinic, and multidiffusive instabilities). As the turbulence acts to reduce its cause (i.e., the angular velocity gradients), it can be modelled as a diffusive process which transports both angular momentum and chemicals together with meridional circulation \citep{Chaboyer1992}.  
In Zahn's framework, the main sources of turbulence are  vertical and horizontal shears. 
Several prescriptions have been proposed for the corresponding diffusion coefficients, some based on multi-D numerical simulations 
\citep[][and references therein]{2022csss.confE..92P}. 
 
Those must be coupled self-consistently to the treatment of meridional circulation as an advective process in stellar evolution models (the advection term for meridional circulation is very important, as it allows angular momentum to be transported up the gradient of angular velocity $\Omega$). 
The resulting chemical mixing properties are directly related to the angular velocity gradients, which also depend on how angular momentum is extracted from the stellar surface, depending on its \teff.

A and F-type dwarf stars with \teff ~above $\sim$~6800~K 
are not efficiently spun down by a magnetic torque (see $\upsilon sini$ in Fig.\ref{LiBeBdipVsini}), as their convective envelope is either inexistent or too shallow for magnetic generation via a dynamo process. A small flux of angular momentum resulting from the respective contraction and expansion of their core and surface is however present in their interior, with a stationary regime where shear turbulence and meridional circulation counterbalance each other, and with mean molecular weight gradients below the convective envelope also playing a stabilizing role. Depending on the stellar rotation rate, the resulting weak mixing can  counteract the effects of atomic diffusion on metals \citep{2006ApJ...645..634T,2020A&A...633A..23D}, while preserving the photospheric Li along the main sequence. This is supported by the fact that Am stars, which are relatively slower rotators than normal A stars, are Li-deficient compare to A stars (by a factor of $\sim$3), and show uniform over- and underabundances of metals \citep{2000A&A...354..216B}. 

The convective envelope of dwarf stars becomes deeper through the \teff \ range between $\sim$ 6900 and 6500~K. The surface dynamo can sustain a  magnetic torque which spins down the outer stellar layers (Fig.\ref{LiBeBdipVsini}). The resulting angular momentum extraction induces an increase of the angular rotation gradients between the stellar core and surface, 
resulting in stronger  meridional circulation and shear turbulence and associated  mixing. As a consequence, stars with cooler \teff ~are expected to experience  larger destruction of both Li and Be, as observed in the dip. Early  Type I rotating models that use specific prescriptions for vertical and horizontal  shears \citep[][respectively]{1997A&A...317..749T,Zahn1992} 
are able to very well reproduce the blue side of the Li and Be dips as it deepens in open clusters of increasing age \citep{1998A&A...335..959T,1999A&A...351..635C,2010A&A...510A..50S}.

Finally, the convective envelope is deeper for G-type dwarf stars on the red side of the Li and Be dips (\teff ~lower than $\sim$ 6500~K), resulting in stronger magnetic torques that spin down the outer layers even more efficiently compared to their hotter counterparts 
(Fig.~\ref{LiBeBdipVsini}). Consequently, the above-mentioned Type~I rotating models that fit well the blue side of the dip  predict too much Li and Be depletion on the cool side of the dip and in G-dwarfs. Conversely, using different prescriptions for the vertical and horizontal shears  
\citep[i.e.][respectively]{Zahn1992,Mathis2018} 
allows for reproducing the cool side of the dip, but not the hot side \citep{Dumont2021b}. This emphasises that one of the major limitations of the treatment of rotation in stellar models comes from the uncertainties on the source and magnitude of shear turbulence. 
A similar difficulty was emphasised by other studies using different prescriptions for rotation-induced mixing \citep{2016A&A...590A..94C}, 
or interactions between helium settling and rotation-induced mixing 
\citep{2003ApJ...587..795T}; in the later case, one of the two free parameters describing the mixing has to be modified by more than one order of magnitude to fit both the hot and the cool side of the dip. 

These results support the conclusion by \citet{1998A&A...335..959T} that the Li dip corresponds to a transition region where another mechanism participates to the transport of angular momentum together with meridional circulation and shear turbulence, resulting in a different efficiency of the associated  mixing of chemicals in dwarf stars of different \teff ~(Sect.~\ref{TypeIImodelswaves}).  
This conclusion is also directly linked to the main caveat of all Type~I rotating models, which predict that the core of solar-type stars would rotate much faster than their surface \citep{
1995ApJ...441..865C,1997PhDT........18T,
2005A&A...440L...9E,2010ApJ...715.1539T,2013A&A...549A..74M}. This is ruled out by helioseismology,  which revealed that the radiative interior of the Sun rotates almost as a solid-body down to about 0.2 solar
radii 
\citep[][and references therein]{2007Sci...316.1591G}. 
Such asteroseismic precision is not  reachable for main sequence stars other than the Sun.  
However, the degree of differential rotation between the deep layers and the surface could be evaluated for a sample of main sequence stars with masses below $\sim$1.5\ms ~observed with {\it{Kepler}} \citep{2015MNRAS.452.2654B,2018Sci...361.1231B,
2015A&A...582A..10N}, with  
 strong indications that rotation at the surface and in the interior are generally close to each other (within a factor of $\sim$2), at odds with Type~I models predictions. 
 
Considerable theoretical effort has been dedicated to identifying the still-missing transport process(es) for angular momentum \citep[][and references therein]{2019ARA&A..57...35A}. We focus on the studies that are directly related to the possible consequences for Li depletion in low-mass stars, and call those models Type~II rotating models.

\subsection{Type II rotating models with  magnetic instabilities} 
\label{TypeIImodelsmagnetic}

Large-scale fossil magnetic fields were  invoked to account for the  rotation profile in the solar radiative interior \citep{1987MNRAS.226..123M,1993ApJ...417..762C,
1999ApJ...511..466B}. As of today, the difficulty is to simultaneously account 
for the latitudinal differential rotation in the convective envelope and  the thickness of the transition region, known as the tachocline  \citep{1992A&A...265..106S,1998Natur.394..755G,2005EAS....17..157Z}, 
with contradictory insights from different numerical simulations 
\citep[for a review see][] {2023SSRv..219...87S}.

The impact of magnetic instabilities on the internal transport of angular momentum is an interesting  alternative,  with much focus on the Tayler-Spruit instability \citep[hereafter TSI;][]{1973MNRAS.161..365T,1999A&A...349..189S,2002A&A...381..923S}. 
A small-scale magnetic field may be produced
in stably stratified layers in stars through this instability, given that the initial magnetic field of the star is small enough and that the 
 differential rotation, which is invoked as the energy source of the process, reaches a minimum degree \citep[for the latest analytical instability criteria, which is a key issue for a self-consistent modelling of AM transport with the TS
dynamo, see][and references therein]{2025ApJ...988..195S}. Numerical simulations, which are not yet  performed under realistic stellar conditions, provide contrasted results \citep{2017RSOS....460271B,2024A&A...681A..75P}. The saturation of the instability and the
resulting angular momentum  transport thus remain poorly understood. 

In stellar evolution models, the vertical transport of angular momentum by the 
TSI in the radiative layers is simulated through the introduction  of a magnetic viscosity (i.e., a parametrized diffusion coefficient) in the equation for the transport of angular momentum; it comes in addition to the terms describing the transport of angular momentum by meridional circulation and shear turbulence considered in Type~I rotation models.  
The main issue  
is that different ad hoc parametric prescriptions are required to reproduce the angular velocity gradients at different phases of the evolution of low-mass stars, i.e., the main sequence,  
the subgiant phase, 
and the red giant phases during hydrogen
shell burning 
and core helium burning \citep{2012A&A...540A.143M,2024A&A...681L..20M,2014A&A...564A..27D,2014ApJ...788...93C,2016ApJ...817...65D,2017A&A...602A..62T,2018A&A...616A..24G,2019MNRAS.485.3661F,2022A&A...664L..16E,2022A&A...663A.180M,2024A&A...689A.307B}, 
with no clear justification as to why the instability would change from one regime to another. 
Additionally, the measurement of the rotation of the deep solar core below 0.2~R$_{\odot}$ using gravity modes is still required 
to legitimate the egibility of the TSI. 

The TSI itself is not expected to transport the chemicals efficiently \citep{2004A&A...422..225M}. Indeed, while
the  transport of AM is ensured by the magnetic stresses, the transport of chemicals results from the motions of the fluid, more explicitly 
by the shear and meridional circulation as in Type~I rotating models. However,   those mechanisms are themselves regulated by the magnetic
instability to the critical value of shear above which it is triggered. As a consequence, a more efficient TSI leads to lower angular velocity gradients, hence to 
lower mixing efficiency. 
However, the different formulations for the TSI lead 
either to the correct depletion of lithium, but resulting in excessive depletion of Be in the Sun, or to no depletion of Li and Be at all \citep{2025A&A...694A.285B}. 
Quoting these authors, this {\it{``casts doubt on the ability to link the observed light-element depletion to the combined effects of shear, circulation, and magnetic
Tayler instability"}}. To the best of our knowledge, no attempt has been made to reproduce the Li-Teff-age patterns with this mechanism, which anyway does not pass the solar Li and Be test.  

Magneto-rotational instabilities, and in particular the azimutal magnetorotational instability (AMRI), which is a  
destabilization of hydrodynamically stable differential rotation by current-free toroidal magnetic fields, is an  alternative that might be favoured.  
It extracts its energy from  differential rotation, which makes it more efficient for steeper rotation
gradients, and ineffective in the case of solid-body rotation. The corresponding turbulent viscosity 
is expected to be very effective at transporting angular momentum  (but much less at transporting chemicals) in a time-dependent way that can explain the observed post-main sequence rotational evolution of
low-mass stars \citep{2015A&A...573A..80R,2016A&A...589A..23S,2020A&A...641A..13J,2024A&A...683A..12M}.
However, using the expression for the parametric viscosity proposed by \citet{2016A&A...589A..23S} that depends on the radial rotational shear and which is expected to be consistent with the AM transport by the AMRI, \citet{2023A&A...677A.119D} showed that Li and Be of main sequence stars can't be reproduced simultaneously. 
This  calls into question the actual relationship between the AMRI instability and the Li and Be depletion history in solar-type stars. 

Whether a magnetic instability could  account for the Li abundance in dwarf stars of different masses and metallicities remains to be investigated, but the current prescriptions do not pass the Li and Be test in the Sun. The Li dip could  once again be the ultimate test, with the pending issues of the triggering of the different magnetic instabilities and the related prescriptions to be included in stellar evolution models, both awaiting for multi-D numerical  simulations performed under realistic stellar conditions. 

\subsection{Type II rotating models with  magneto-inertial gravity waves} 
\label{TypeIImodelswaves}

{\it{``The results of this paper, if correct, may be an unwelcome addition to the theory of stellar interiors" \citep[][``Radiative and other effects from internal waves in solar and stellar interiors"]{1981ApJ...245..286P}.}} 

\subsubsection{The first  models including rotation and internal gravity waves}
\label{TypeII-IGW}

As with atomic diffusion and rotation-induced processes, the study of the Li dip has highlighted the need to investigate the role of internal gravity waves (IGW) in low-mass stars. As explained in Sect.\ref{TypeImodels}, this  feature falls in a range in effective temperature that corresponds to a transition region for the structural properties of dwarf stars (in particular, the development of the H-He convective envelope) that impact both their surface rotation and  Li  abundance. It was also stated that it is a transition region where a very efficient transport of angular momentum surpasses meridional circulation and turbulence  \citep{1998A&A...335..959T}. This turned out to be  characteristics of IGW. Using a simplified formalism for bulk wave excitation through Reynolds stresses (more details below),  
it was shown that the growth of the convective envelope in dwarf stars of both Pop~I and Pop~II with \teff ~below $\sim$~6600~K  coincides to the growth of the generation and efficiency of IGW in their interior  \citep{2003A&A...405.1025T,Talon2004}.
After the introduction of this excitation formalism together with a semi-analytical framework to treat wave damping and the associated transport of angular momentum together with that due to meridional circulation and shear in stellar evolution codes \citep{2005A&A...440..981T}, the results were extremely encouraging. Not only did the models succeed in achieving near uniform rotation in the Sun and solar-type stars, 
but they could reproduce the behaviour of Li in both Pop~I and Pop~II dwarf stars, including the Li and Be dips and the flatness of the Li plateau \citep{2005A&A...440..981T,2010IAUS..268..365T,2005Sci...309.2189C,2005EAS....17..167C,2010IAUS..268..341M}. 

\begin{figure}
\begin{center}
{\includegraphics[angle=0,width=0.17\textwidth]{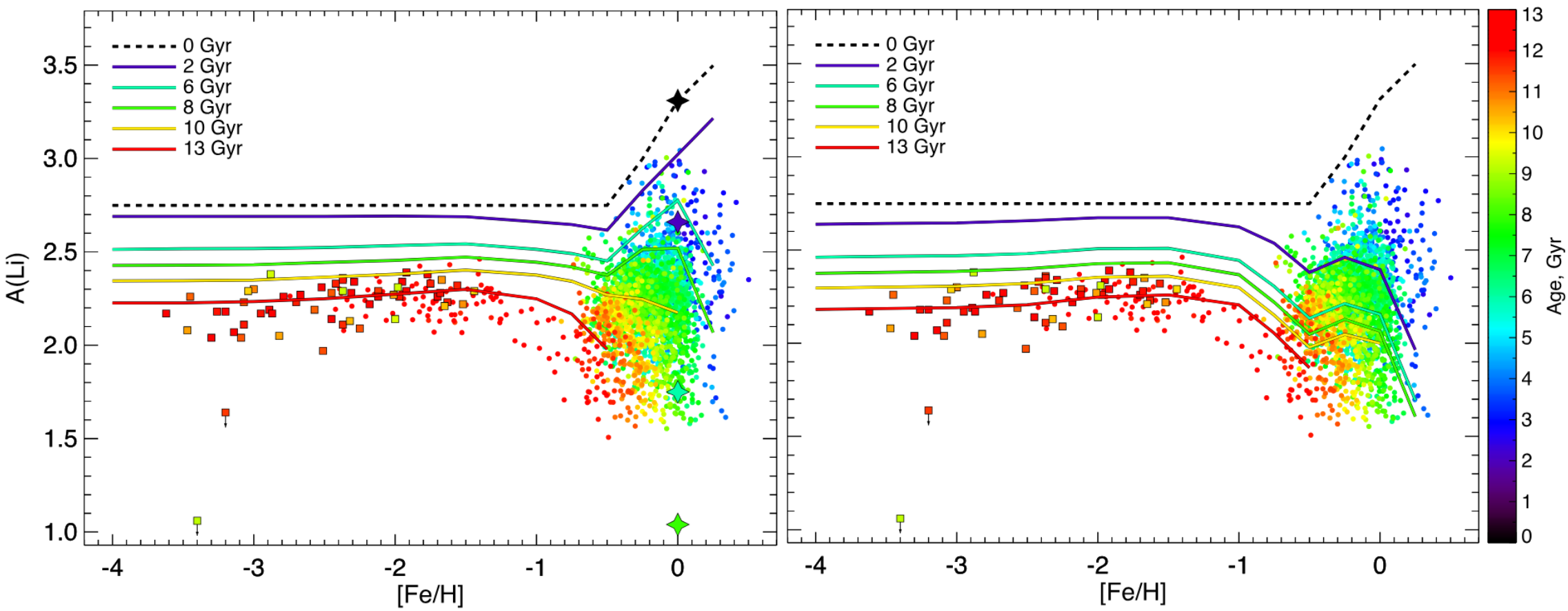}}
\end{center}
\vspace{0pt}
\caption{ 
Li abundance (3D, NLTE) as a function of \afe \ for the \citet{Norris_2023}
and GALAH DR3 sample stars (squares and circles respectively) with colour-coded
age. Solid lines show the theoretical Li upper and lower
envelopes (left and right panels, respectively) at different ages (in the legend), i.e., the maximum and minimum Li values at a given [Fe/H]
and age expected from Type~II models with 5800 K$<$ T$_{\rm eff}$(ZAMS) $<$6500 K. The dashed line indicates the initial Li abundance assumed in the models. In the left panel, the four-pointed stars
show predictions for 1 \ms \ star from models of  \citet{Dumont2021a}  
with age similarly colour-coded. Figure from \citet{Borisov2024}
 }
 \label{fig_Li_depletion}
\end{figure}

\subsubsection{Towards models including rotation, internal gravity waves, magnetism, and their interactions}
\label{TypeII-MGWI}
If all seemed to be working self-consistently so well when considering rotation-induced processes and internal gravity waves, why weren't these results the end of the quest to understand Li behaviour in low-mass dwarf stars? 
First, this series of Type~II models with IGW were computed with the only  combination of prescriptions for vertical and horizontal shears available at that time \citep[][respectively]{1997A&A...317..749T,Zahn1992}.  
This resulted in a perfect shaping 
of the Li dip  
while explaining the solar Li. 
\citet{Dumont2021b,Dumont2021a} showed, however, that this  excellent agreement didn't hold anymore when new   prescriptions for the vertical and horizontal shears (\citealt{Zahn1992} and \citealt{Mathis2018} on one hand, and \citealt{Zahn1992} and \citealt{Mathis2004} on the other hand) as validated by direct numerical simulations \citep{2013A&A...551L...3P,2017ApJ...837..133G}, 
were included in Type~II rotating models.  
In \citet{Dumont2021b,Dumont2021a}, 
 a parametric viscosity (including \citealt{Spada2016}'s dependence of the angular momentum transport efficiency on the radial rotational shear) 
 was added to meridional circulation and shear to simulate the missing AM transport mechanism; mixing was also parametrised to fit Li 
 in cool dwarfs, including the Sun and solar-type stars. 
 
 Interestingly, the same models  simultaneously account for the Li plateau \citep{Borisov2024} as shown in Fig.~\ref{fig_Li_depletion}, offering for the first time a coherent view of Li depletion and angular momentum transport from Pop~~I to Pop~II dwarf stars. 
However, 
the new shear prescriptions fail to explain the Li dip even
when combined with additional sources for turbulent transport. Consistently with \citet{2010IAUS..268..365T}'s  early findings, the only Type~II rotating models of \citet{Dumont2021b} that could reproduce simultaneously the Li dip and the Li behaviour in G-type stars are those including the oldest prescriptions for vertical and horizontal shear turbulence, together with a mass-dependent efficiency for the additional AM transport process, as expected from IGW. Quoting these authors, {\it{``Insight and validation of the different available prescriptions for (...) turbulent shear transport from hydrodynamicists and
multi-dimensional numerical simulations (...) 
are now mandatory in order to overcome
this impasse."}}
 
The second reason to revisit Type~II models including IGW is that  more sophisticated treatment for IGW and their interactions with rotation and magnetic field must  be taken into account. IGW  naturally occur 
and propagate in fluids that are stably stratified, with gravity acting as the restoring force. 
Low-frequency progressive waves have long been considered to play an important role in the redistribution of angular momentum in low-mass stars spun down by a magnetic torque 
\citep{1981ApJ...245..286P,
1993A&A...279..431S,1997A&A...322..320Z,1997ApJ...475L.143K,2002ApJ...574L.175T}.  
They might also trigger chemical mixing \citep{1991ApJ...377..268G,1994A&A...281..421M,
2000A&A...354..943M,2003ApJ...595.1114Y,
2017ApJ...848L...1R}.  
However, this direct wave-induced mixing was usually neglected in Type~II rotating models including IGW, where the transport of chemicals is still ensured by the meridional circulation and the shear regulated by the impact of the IGW on the angular velocity gradients \citep{2005A&A...440..981T}.
Even more importantly, IGW carry angular momentum from the region where they are excited to the region where they are dissipated. 
A deep understanding of their excitation, propagation, and  dissipation mechanisms is thus of key importance.  

Third, many studies have focused on the generation of IGW by turbulent convection at the interface between convective and radiative layers in stars (in low-mass stars, the excitation is related to the  convective envelope),  neglecting the impact of rotation and magnetic fields on wave excitation (here we do not consider tidal excitation but focus on single stars). 
IGW are excited by two different processes, namely, turbulent pressure in the convective bulk,  
which is predominant in solar-type stars,   
and penetrative convection by so-called convective plumes at
the interface between the convective and the radiative zones 
\citep[for a review see][]{2015EAS....73..111S}. 
The existence of those waves
is observed in 2 and
3–D numerical simulations  \citep{1986ApJ...311..563H,1994SoPh..152..241A,2000AcA....50...93K,2005A&A...438..365D,2005MNRAS.364.1135R,2014A&A...565A..42A,2023MNRAS.522.2835L},   
but the extrapolation   
to the stellar regime remains a 
challenging task.  Consequently,  semianalytical prescriptions for excitation have been  necessary in a first approach, but they provide only crude descriptions of the generation of IGW in stars. Additionally, the prescriptions that have been used so far in Type~II stellar evolution models neglect the action of rotation and magnetism on turbulent convection and therefore on stochastic wave excitation, as studied in numerical and laboratory experiments 
and theoretical developments \citep[][and references therein]{2024A&A...690A.270B}. 
Recently, theoretical progress was achieved to modify the MLT convection formalism to account for rotation and magnetic field  \citep[][Rotating and Magnetised Mixing-Length Theory]{2025arXiv250514650B} as well as the formalism for stochastic excitation with both rotation and magnetic field on acoustic modes (Bessila \& Mathis, private communication), as required to explain the non detection of acoustic waves in about half of the solar-type stars observed with {\it{Kepler}} \citep{2019FrASS...6...46M}. Extending these formalisms to the case of gravity waves, and testing them in a new generation of Type~II stellar evolution models, is a priority.   

The next theoretical challenge is to establish realistic prescriptions for the propagation and the dissipation of the waves in the radiative zones, which drive the  extraction or  deposition of angular momentum in stellar interiors. The first Type~II evolution models including angular momentum  transport by IGW assumed that radiative damping is the dominant process 
\citep{2005A&A...440..981T,2014ApJ...796...17F}. However, the role of critical layers (when the frequency of the wave is the same as the local rotation frequency; \citealt{2013A&A...553A..86A}) and of nonlinear breaking 
(due to the overturning of the stratification, i.e., convective instability or to the 
shear of the wave itself, i.e., Kelvin–Helmholtz instability; \citealt{2010MNRAS.404.1849B,2025A&A...694A.173M}) have been neglected so far. 

Last but not least, low-frequency IGW that transport angular momentum can be modified by the Coriolis acceleration and the Lorentz force. This results in the IGW becoming gravito-inertial \citep{2007A&A...474..155P,2009A&A...506..811M,2016JFM...800..213M,2018A&A...615A.106P} or magneto-gravito-inertial \citep{1999JFM...398..271D,
2010MNRAS.401..191R,2011A&A...526A..65M,
2012A&A...540A..37M,2017MNRAS.466.2181L}. 
 
It is imperative that efforts are pursued to describe the related physics and to develop reliable prescriptions to be included into stellar evolution models. We have no doubt that both the internal rotation revealed by asteroseismology in various eras of the HRD (using the legacy of {\it{Kepler}}/K2, Tess, and the future PLATO) and the Li abundance patterns (using the legacy of large spectroscopic surveys) will provide the best constraints to this endeavour.
We are confident that the self-consistent treatment of rotation, internal gravity waves, and magnetic field, whose respective impacts have been estimated through simplified prescriptions in already complex stellar evolution models, will confirm that these inter-dependent mechanisms are responsible for the Li-Teff-age-metallicity abundance patterns in low-mass dwarfs.

\subsection{Lithium depletion along the red giant branch} 
\label{theoryadvancedphases}

{\it{Giants with ``(...) Li abundances in agreement with
the predictions of standard models are the exception" \citep{2016A&A...591A..62S}.}}
Not surprisingly, 
the classical and standard stellar  evolution models which fail to reproduce the main sequence Li data also fail to explain the Li abundances observed in post-MS low-mass stars (Sect.~\ref{Li-depletion-obs-evolved}). 
In the case of stars 
originating from the left part of the MS Li dip and belonging both to the field and to OC, 
Li drops immediately after the turnoff and before the theoretical occurrence of the 1DUP \citep{2003A&A...409..251M,2004A&A...424..951P,2014ApJ...785...94L,2019AJ....158..163D,2020A&A...633A..34C}.  
This is related to the fact that the weak mixing that counterbalances atomic diffusion while those stars are on the main sequence enlarges their Li-free region compare to classical models \citep{2003A&A...399..603P,2004A&A...424..951P,2010A&A...522A..10C}. 
Additionally, many post-1DUP giants have Li upper limits, which can be explained only if they have undergone Li depletion on the main sequence, in agreement with the main sequence Li data, with rotation-induced mixing being favoured \citep{2025arXiv250704266S}. 
We emphasize that main sequence Li depletion followed by the 1DUP are sufficient to explain the Li depletion observed in stars crossing the Hertzsprung gap and climbing the lower part of the RGB \citep[][and references therein]{
2020A&A...633A..34C,
2021A&A...651A..84M}, as well as that of Be \citep{2010A&A...510A..50S}. No extra-mixing leading to additional Li destruction is required during these phases. 
This brings important constraints to the angular momentum transport processes in post-MS stars. 
Recent
asteroseismic observations based on space photometry have uncovered the internal rotation of low-mass subgiants, 
red giants 
during hydrogen
shell burning, 
and clump stars during helium burning 
unambiguously showing that their cores rotate order of magnitudes slower than predicted by Type~I rotation models \citep[][and references therein]{2020A&A...641A.117D,2018A&A...616A..24G,2024A&A...681L..20M}. Based on semi-parametric models, it has been shown that during the subgiant phase, the 
 angular momentum transport
efficiency increases with the mass of the star and decreases as the star evolves to the bottom of the RGB, and that its efficiency  
increases with the mass of the star as well as along the climbing of the RGB \citep{2016A&A...589A..23S,
2022A&A...663A.180M,2023A&A...677A.119D}. 
Whatever the origin of the angular momentum transport is, it should not affect the photospheric Li abundance in these stars. 

Later on along the RGB, however, the
additional and sudden photospheric Li drop observed together with a decrease of $^{12}$C/$^{13}$C and C/N at the so-called luminosity bump requires the occurrence of a very efficient hydrodynamical instability to transport Li from the base of the convective envelope to its burning region in the outer wing of the hydrogen burning shell where the CN-chain is operating.  
As of today, the only plausible mechanism to self-consistently explain these observations is the so-called thermohaline instability which is expected to set in at the luminosity of the RGB bump for red giant stars of low-mass  that ignite He-burning in their degenerate core through so-called He flashes\footnote{In more massive stars, this process should not occur on the RGB - since core He-burning starts in non degenerate conditions before the stars could reach the  bump - 
but on the early-AGB.} \citep{2007A&A...467L..15C,
2019A&A...621A..24L,
2023A&A...670A..73A}, as corroborated by the seismic analysis of bright red giants observed by CoRoT \citep{2015A&A...580A.141L} and {\it{Kepler}} \citep{2024MNRAS.527.8535C}. Despite its successes to also explain the Galactic evolution of $^3$He \citep{2012A&A...542A..62L,2022ApJ...936..168B}, the details of the thermohaline mixing are debated \citep{2011A&A...533A.139W,2009ApJ...696.1823D,
2011ApJ...727L...8D,
2022ApJ...935L..30T}, and some observations for red giants belonging to globular clusters are found to deviate from theoretical prescriptions \citep{2015MNRAS.450.2423A,2017MNRAS.469.4600H,2023MNRAS.524.4418M}. Several attempts to simulate numerically thermohaline convection have led to contradictory results \citep{2010ApJ...723..563D,
2011ApJ...728L..29T,2013ApJ...768...34B}. Interestingly, it was found that rotation and/or weak magnetic fields can strongly enhance the vertical transport by this instability, at the level that is necessary to explain the
Li behaviour and the concurrent carbon isotopic and C/N ratios changes observed in RGB stars around the luminosity bump 
\citep{2018ApJ...862..136S,2019ApJ...870L...5H,2024ApJ...964..184F}.

\section{PRODUCTION OF LITHIUM IN ASTROPHYSICAL ENVIRONMENTS}
\label{sec:LiProD}

In stellar interiors, the minor lithium isotope \lia \ can be produced directly by thermonuclear reactions between stable isotopes through the reaction
$ \rm ^9Be + p \longrightarrow  ^6Li + ^4He $,   
\noindent but 
negligible amounts of it may survive in any environment. The direct production of the major isotope \lib \ involves the unstable isotope $^3$H (tritium, lifetime of 12.5 yr) through
$\rm ^4He + ^3H \longrightarrow  ^7Li$. 
\noindent 
Tritium does not exist in stellar interiors but it is formed in the conditions of primordial nucleosynthesis from the capture of the abundant neutrons by deuterium and participates in the formation of \lib \  in that environment (\S~ \ref{subsec_bigbang}).

The main mechanism of \lib \ formation in stars is through the decay of the radioactive isotope \bea \, which is produced in various stellar environments via
$\rm ^3He + ^4He     \longrightarrow ^7Be$.
In normal conditions \bea \ decays through electron capture of a K-shell electron with a lifetime of 53.3 days. \cite{Cameron_1955} noticed that in the hot stellar interiors the half-life of \bea  \ is lengthened nearly inversely proportionally to the degree of ionization of its K-shell electrons: at temperatures of 10$^7$-10$^8$ K its lifetime may be extended to tens of years, allowing \bea \ to be transported by ``circulation currents" from the hot regions to the surface. He argued that this mechanism could conciliate the observations of giant stars showing simultaneously low $\rm ^{12}C/^{13}C$ ratios (a signature of CN cycle) and high Li. This mechanism was reinvestigated in \cite{Cameron1971} who extended it as to account for the simultaneous presence in S-stars of both Li and s-elements (made in the He-shell of AGB stars). \footnote{\cite{Cameron1971} acknowledge that the Li transport in their scheme follows  \cite{Cameron_1955}. It should be then called - when applied specifically to the case of Li -  the "Cameron mechanism" and not the "Cameron-Fowler mechanism" as usually done. }

\begin{figure}
\begin{center}
{\includegraphics[angle=270,width=0.7\textwidth]{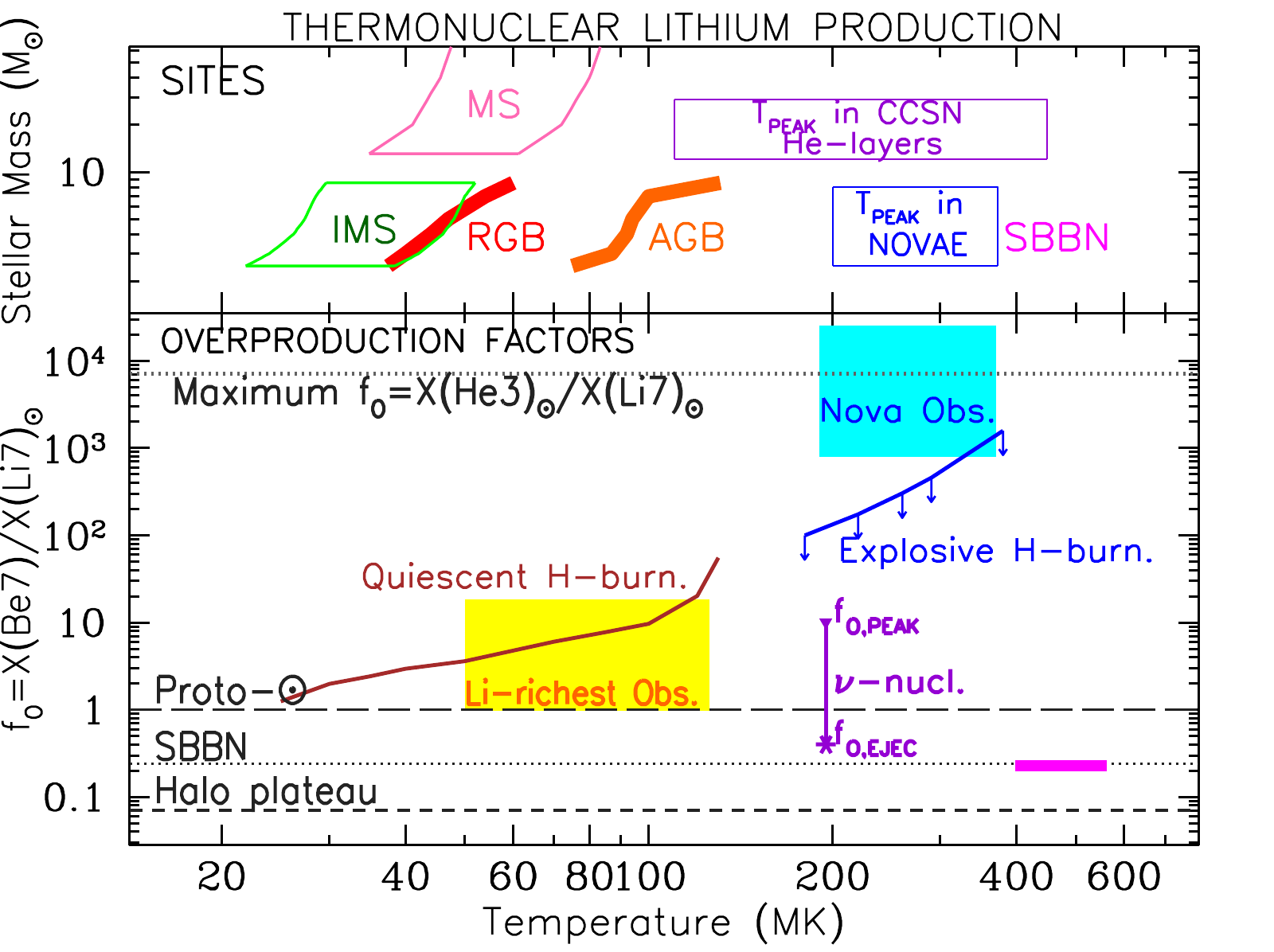}}
\end{center}
\vspace{30pt}
\caption{ 
{\it Top}: Stellar astrophysical sites of thermonuclear production of \bea \  with corresponding temperature ranges: core H-burning (horizontally extended from H-ignition on the left to H-exhaustion on the right),  in Intermediate Mass Stars (IMS) and Massive Stars (MS), shell H-burning in  Red Giant Branch stars (RGB, red) and Hot Bottom Burning in Asymptotic Giant Branch stars (AGB, orange);  range of peak temperatures in He-layers of CCSN of 15 to 30 \ms, where $\nu$-induced nucleosynthesis occurs (violet box), novae (blue box) and  SBBN (purple).\\
{\it Bottom}: Overproduction factors $\rm f_0$ of \bea \ (with respect to the meteoritic value of \lib). The brown solid curve shows results for quiescent H-burning of solar mixture material at constant temperatures. Li is overpoduced, albeit for short timescales: a few 10$^3$ yr for T=25 MK to about a week for T=120 MK. 
The  yellow shaded area indicates observations of Li-richest giant stars. The upper dotted horizontal line indicates the maximum possible value $\rm f_0$, with all solar He$^3$ turned into \bea; some nova observations indicate higher overproduction (see Fig. \ref{fig_Li_BBN_nova}, right).  Blue asterisks indicate maximum $\rm f_0$ values obtained in nova simulations for indicated peak temperatures (above 180 MK) from \cite{Starrfield2024}, while the cyan-coloured shaded area indicates the range from nova observations \citep{Molaro2023}. The vertical violet segment joins peak overproduction f$_{\rm 0,PEAK}$ of \lib \ in the He-layers from $\nu$-induced nucleosynthesis and corresponding value in the total ejecta f$_{\rm 0,EJEC}$ in a 27 \ms \  CCSN model \citep{Sieverding_2019}.  Adopted values (in horizontal lines)  are: A(Li)=3.39 for Proto-solar \citep[long-dashed line]{Lodders_2025}, 2.7 for SBBN \citep[dotted]{Pitrou_2018} and 2.2 for halo plateau \citep[short-dashed]{Bonifacio_2025}.
 }
\label{fig_Li_prod}
\end{figure}

Fig. \ref{fig_Li_prod} presents an overview of  the thermonuclear production of \bea \ in various environments.  In the top panel, typical temperature ranges in stars of various masses and evolutionary phases are displayed.
In the bottom panel, H-burning in stellar cores (main sequence stars), shells (RGBs)  and bottoms of convective envelopes (AGBs) is represented by idealized conditions of constant temperature and density, while for the other cases (novae, BBN, $\nu$-induced nucleosynthesis)  results of recent calculations are shown.  The abundance of \bea \ results from the competition between its production  and its destruction by proton reactions (electron captures being much slower in such temperatures). In the bottom panel it appears that \bea \ may  exist  in  significant amounts for all the shown temperatures, until exhaustion of the $^3$He. The maximum possible mass fraction is the initial mass fraction of $^3$He and this is achieved for temperatures of the order of 200 MK and above (as in nova explosions). This time span is inversely proportional to the density of the site, but the overproduction factor, i.e. its average overabundance w.r.t. proto-solar \lib, is independent of the density. In most cases, the overproduction timespan is much shorter than the evolutionary timescales of stellar interiors and \bea \ is rapidly destroyed,  
unless if it manages to escape to cooler regions through the Cameron mechanism.  
This is the case in both AGB stars and nova explosions, while in the case of BBN and $\nu$-induced nucleosynthesis it is rather the rapid cooling of the site that allows the survival of \bea. Before discussing those sites of thermal production of \bea, we turn into a non-thermal mechanism in the next section.

\subsection{Spallation reactions in cosmic rays}
\label{subsec_CosmicRays}
In view of the fragility of Li, Be and B in stellar interiors, its was  suggested that those elements may be produced by high-energy spallation reactions of the more abundant C, N and O nuclei. 
The acceleration of the spallating nuclei could occur either at the surfaces of flaring solar-type stars \citep{Hayakawa_1955} or of magnetic or T-Tauri star \citep{Fowler_1955, B2FH_1957}, where the low-density environment would preserve their fragile LiBeB products from destruction by thermonuclear reactions. However, \cite{Ryter_1970} showed that the release of gravitational energy powering the activity of those stars would be insufficient to produce the observed amounts of those light nuclei in the Galaxy, because a large fraction of that energy would be spent by the energetic nuclei in ionizing the stellar atmospheres. 

\begin{figure}[h]
\begin{minipage}{.49\textwidth}
{\includegraphics[width=\textwidth]{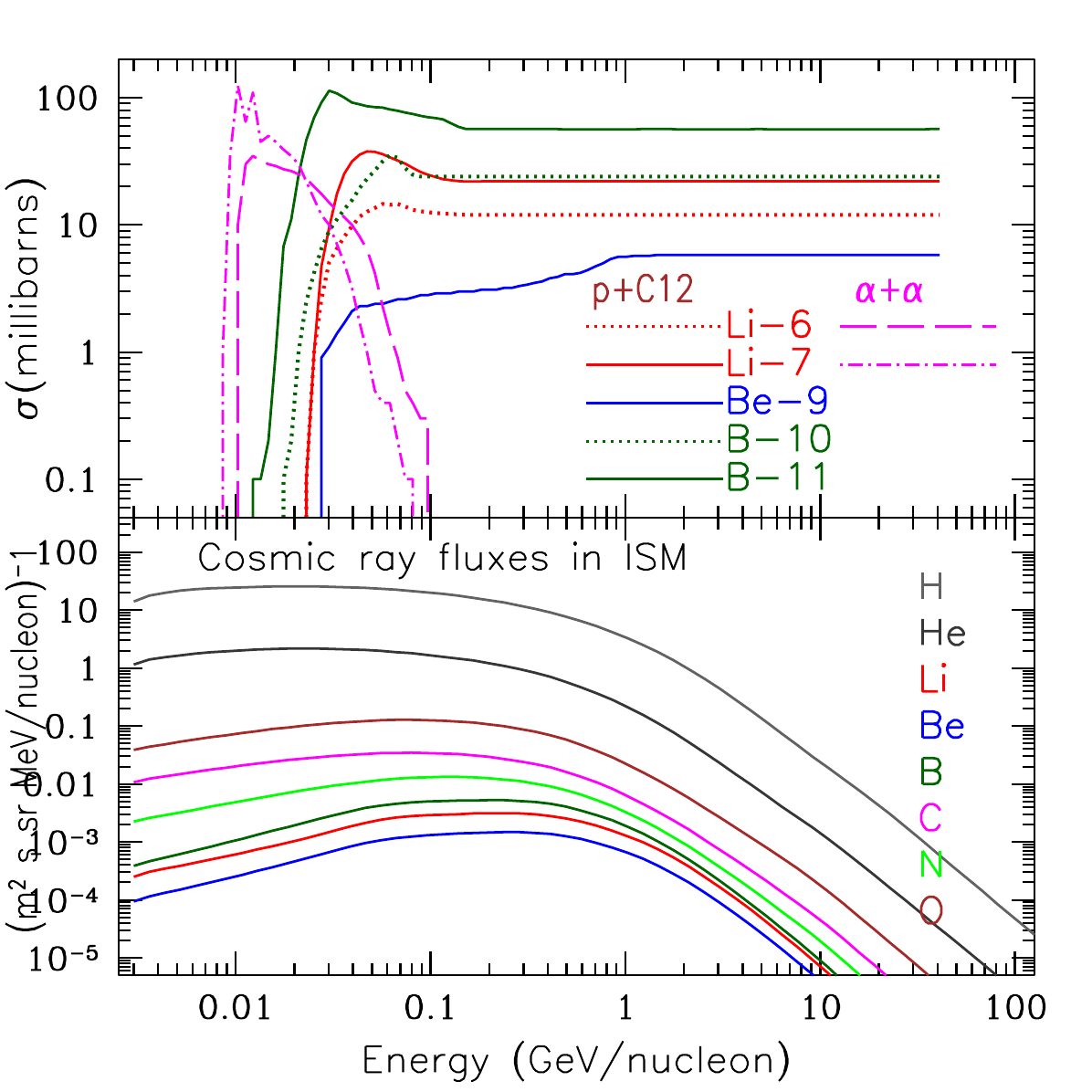}}
\end{minipage}
\hfill    
\begin{minipage}{.49\textwidth}
{\includegraphics[width=\textwidth]{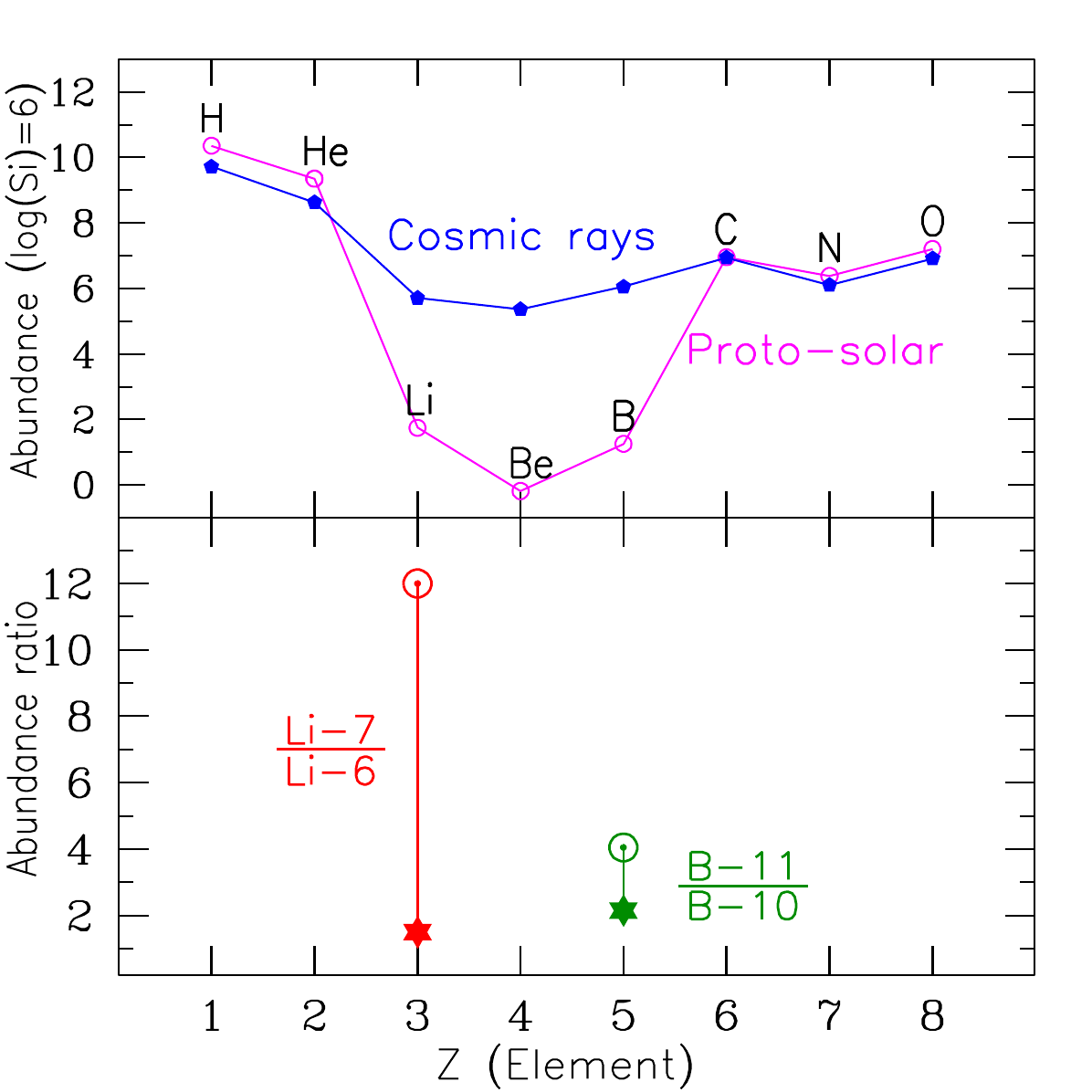}}\label{fig:sub_b}
\end{minipage}
    \caption{{\it Top left:} Cross-sections of the production of $^6$Li, $^7$Li, $^9$Be, $^{10}$B and $^{11}$B through interactions of $^{12}$C  with energetic protons \citep[from][]{Read_1984} and $\alpha+\alpha$ reactions \citep[from][]{Mercer_2001}. {\it Bottom left:} Interstellar Galactic Cosmic Ray (GCR) spectra obtained  by fitting the Voyager 1 observations  with the GALPROP propagation code \citep{Cummings_2016}.
    {\it Top right:} Composition of the proto-solar system \citep[from][]{Lodders_2025} and in GCR \citep[from][]{Cummings_2016}; they are both scaled to 10$^6$ Si atoms. {\it Bottom right:} Abundance ratios in meteorites (solar symbols) and in GCR (asterisks) observed in PAMELA experiment in ISS \citep{Menn_2018} with Li7 representing the sum of detected Li7 and Be7.   }
    \label{fig:GCR_Spectra_Abundances}
\end{figure}

\citet{Bradt_1950} had already noticed that Li, Be and B are present in Galactic Cosmic Rays (GCR) almost at the same  level as the abundant C, N and O nuclei, while the LiBeB/CNO ratio is about a million times lower in proto-solar (meteorite) material (Fig. \ref{fig:GCR_Spectra_Abundances} top right). 
This,
along with the measurements of the production cross-sections of LiBeB from CNO (Fig. \ref{fig:GCR_Spectra_Abundances} top left) measured by \citet{Bernas_1967},
led \cite{Reeves_1970} to suggest that Li, Be and B are produced by spallation reactions between the ISM and GCR, during the propagation of the latter in the Galaxy in the $\sim$10 Gyr of its life. However,  they noticed that the \liba \ ratio produced by cosmic rays is $\sim$8 times lower than in the meteorites (1.5 vs 12.05 respectively), as shown in Fig. \ref{fig:GCR_Spectra_Abundances} (bottom right): in contrast to the other LiBeB isotopes, \lib \ is largely underproduced in GCR.

Subsequent work, taking into account the injection of accelerated cosmic rays from their sources and their propagation in the Galaxy, in the framework of the so-called "leaky-box" model \citep{Meneguzzi_1971} confirmed the conclusions of \cite{Reeves_1970}. A similar, albeit less acute problem, was also identified : the  \bba \ ratio in GCR, obtained through the measured cross-sections, is $\sim$2.2 compared to its meteoritic value of 4.05 (Fig. \ref{fig:GCR_Spectra_Abundances}, bottom right). This difference is much larger than the uncertainties in the production cross-sections or in the meteoritic measurements and, taken at face value, implies that \bb \ requires another, complementary source in order to fully account for its proto-solar abundance.

The mismatch between proto-solar abundances and the GCR production of the Li and B isotopes requested for alternatives.
A low-energy GCR component would favour the production of \lib \ over \lia \ due to the  form of the $\alpha+\alpha$ low-energy cross sections (see Fig. \ref{fig:GCR_Spectra_Abundances}) as suggested by \citet{Meneguzzi_1975}, but it would also enhance the ionization  of the gas in the acceleration site of GCR, which was not found. Moreover, the Voyager 1 data, obtained at the heliospheric border \citep[][and Fig. \ref{fig:GCR_Spectra_Abundances}, bottom left]{Cummings_2016}  indicate that such a component does not exist in the ISM, at least in the local one. The only source suggested so far for \bb \ is neutrino-induced nucleosynthesis in CCSN, which may also contribute to some extent to \lib \ production (\S~\ref{subsec_neutrinos}). 

Taking into account that \lia \ is exclusively produced by GCR, the contribution of GCR nucleosynthesis to the meteoritic (protosolar) Li abundance  (sum of \lia \ and \lib), is $\sim$20\%, i.e. GCR constitute a significant and observationally supported Li source.

\subsection{Primordial nucleosynthesis }
\label{subsec_bigbang}

Following the discovery of the CMB in 1965, primordial nucleosynthesis was investigated by \cite{Peebles_1966} for the H and He isotopes, then by \cite{Wagoner_1967} who considered also  heavier nuclei and found substantial production of \lib. It is produced by both reactions $^4$He+$^3$H$\longrightarrow$\lib \ and
$^4$He+$^3$H$\longrightarrow$\bea.
In the former reaction $^3$H is obtained through D+n$\longrightarrow$$^3$H, due to the availability of a large amount of neutrons (n/p$\sim$1/7) at the ``freeze-out" of charged-current weak interactions. 
The reaction with tritium dominates the production of \lib \ at low densities, while at high densities  \lib \ production is dominated by $^4$He+$^3$H$\longrightarrow$\bea. The abundances of \bea \ and \lib \ depend  on the competition between the nuclear reaction rates (which depend on the baryonic density) and the expansion rate of the Universe. The rapid cooling of the primordial plasma within a timescale of a few minutes ensures naturally the survival of all isotopes of both \lib \ and \bea.

The sensitivity of BBN products (D, $^3$He, $^4$He and \lib) to various parameters of the model - in particular, the baryonic density - makes those isotopes important probes of the physics of the early Universe. 
The discovery of the "Spite plateau" (see \S~ 2) provided a constraint on the physics of the early universe but 
it also suggested that SBBN and GCR together cannot make the totality of proto-solar Li, thus making  another - most probably stellar - source for that element necessary.

Starting with NASA's WMAP in 2003, BBN results and predictions changed both quantitatively and qualitatively. Instead of roughly evaluating the baryon-to-photon ratio from observations of the light element abundances in old/unevolved systems, this ratio is now obtained to high accuracy from the analysis of the CMB,  providing stringent tests of BBN and of physics beyond the Standard Model \citep{Plank_2020}.

The primordial nucleosynthesis results are currently in excellent agreement with  abundance measurements of D in high redshift gas clouds and CMB measurements of baryon/photon ratio \citep[the ``triple point" of D, see][and references therein]{Cook_2024}. However, an important discrepancy persists  between the \lib \ value of the halo plateau and the SBBN predictions. The extent and potential causes of that discrepancy were scrutinized over the years: nuclear reaction rate uncertainties \citep{Iliadis_2020}; observational determination of Li abundance in field halo stars or globular clusters \citep{Boesgaard_2024}, particle physics beyond the Standard Model in the early universe \citep[][and references therein]{Bertulani_2023}, 
or destruction of Li in the atmospheres of halo stars. In \S~ \ref{Depletion-obs-PopIIhalodwarfs} we argued for the latter solution to the cosmological problem of Li. Recently, {\citet{Molaro_2024} reported the observation  of A(Li)$\sim$2.2, i.e. the halo plateau, in the Small Magellanic Cloud's gas; if confirmed, this might put in question the stellar solution to the cosmological Li problem.

The cosmologically inferred value of primordial Li, A(Li)=$\sim$2.7,  suggests that primordial nucleosynthesis has a confirmed and significant contribution to the protosolar Li abundance of A(Li)=3.39 \citep{Lodders_2025}, of the order of $\sim$20\%, which is similar to the one of GCR. This implies that the largest part of protosolar Li, $\sim$60\%, is produced by stars.

\subsection{Hot Bottom Burning in Asymptotic Giant Branch stars}
\label{subsec:HBB_AGB}

Despite its fragility, \lib \ can be produced as \bea \ in 
intermediate-mass stars during the thermal pulse phase on the Asymptotic Giant Branch (TP-AGB) through Hot Bottom Burning in the base of the convective envelope \citep{Scalo_1975}. 
 The preservation of $^7$Li requires the introduction in the stellar evolution models of a suited time-dependent convective diffusion algorithm for the transport of chemicals in the convective envelope.
\begin{marginnote}
\entry{Hot Bottom Burning}{Stars with mass $>$3-4 \ms \  develop very high temperatures ($>$40 MK) at the base of their deep convective envelopes, allowing for nuclear burning. }
\end{marginnote}
$^7$Li can be produced efficiently as long as $^3$He is abundant and the HBB proceeds, depending on stellar mass, metallicity and the adopted mass loss rate. Later in the evolution, 
the surface $^7$Li abundance drops again, either because the HBB stops due to mass loss reducing the mass of the envelope, or because
$^3$He is almost completely depleted in the envelope and  $^7$Li is is quickly destroyed by proton capture. 
It is thus expected that Li-rich TP-AGB stars appear in a limited luminosity domain which depends on the aforementioned factors \citep{1997A&AS..123..241F,1999A&A...348..846M,2000A&A...363..605V,
Travaglio2001,Karakas2016,Ventura_2018}. The models are constrained by the observations of Li-rich TP-AGB stars in the Galaxy and in the Magellanic Clouds \citep{Smith1989,
1990ApJ...361L..69S,1991A&A...245L...1A,1995ApJ...441..735S,2007A&A...462..711G}. 

The short duration of the Li-rich phase in TP-AGB stars has cast doubt on their ability to contribute significantly to $^7$Li enrichment on a galactic scale \citep{Romano2001}, in contrast to earlier expectations \citep{Abia1993}. The complexity of those phases, involving poorly understood mechanisms of mixing and mass loss,  
make the robust prediction of Li yields  difficult \citep{Cristallo_2020,Choplin_2024}. 
The predicted net Li yields of models with sub-solar metallicities are, in general, quite small or even negative, i.e., Li is destroyed rather than produced. Positive net yields are obtained in the calculations of \citet{Karakas2016} for stars of twice solar metallicity (Fig. \ref{fig_Li_BBN_nova}, left) and, interestingly,  they are just 
at the level required by GCE models to produce most of the proto-solar Li. Taken at face value, the AGB yields of \citet{Karakas2016} for metallicities up to solar fail to  reproduce the proto-solar Li abundance by large factors. However, in view of the many uncertainties in the treatment of this complex phase of stellar nucleosynthesis it would be premature to exclude AGBs as an important Li source. 

\begin{marginnote}
\entry{Stellar Yield}{Total amount of a given element ejected by a star of a given mass. }
\end{marginnote}

\begin{marginnote}
\entry{Net yield} {The difference between the ejected mass of an element and the one that originally entered the ejected part of the star. 
}
\end{marginnote}

\begin{figure}[h]
\begin{minipage}{.49\textwidth}
{\includegraphics[width=\textwidth]{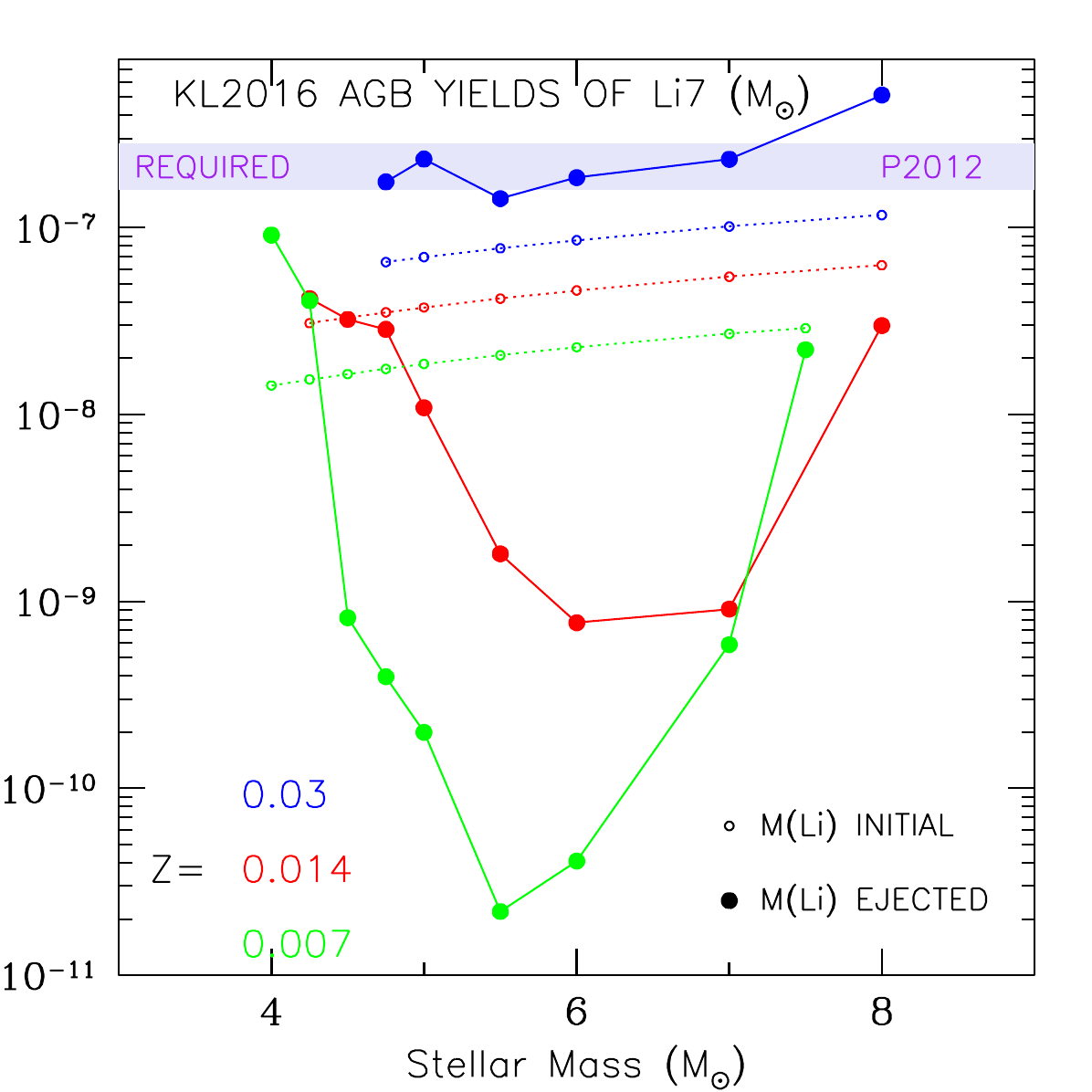}}
\end{minipage}
\hfill    
\begin{minipage}{.49\textwidth}
{\includegraphics[width=\textwidth]{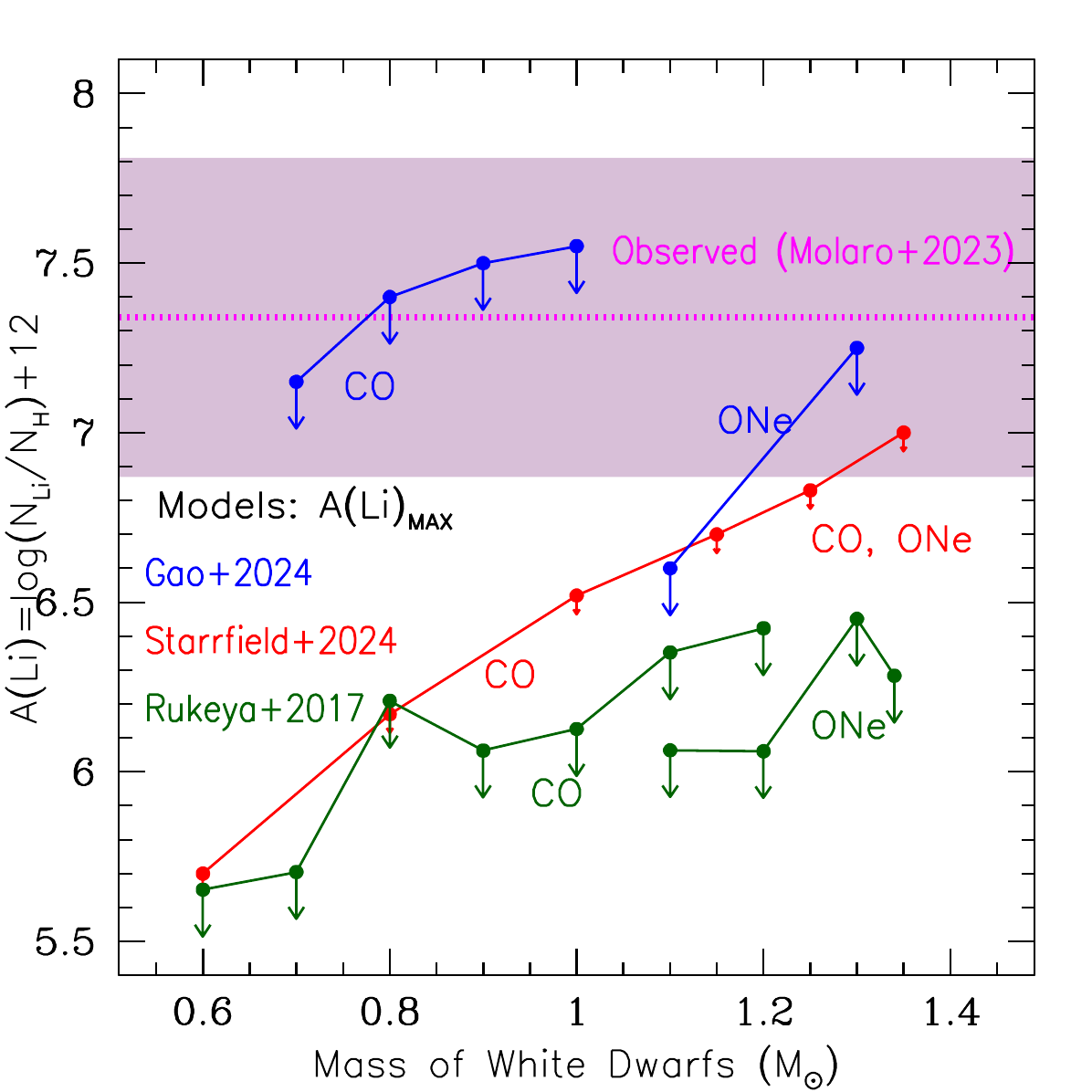}}
\end{minipage}
\caption{{\it Left}: AGB yields of Li from HBB in models of stars of 4-8 \ms \ from \citet{Karakas2016}, for 3 initial metallicities. Ejected masses (solid lines) are lower than initial ones (dotted)  and insufficient to produce proto-solar Li \citep[as  indicated by the shaded aerea, obtained through GCE models in][]{Prantzos2012}, except for supersolar metallicity stars (blue solid). \\
{\it Right:} Relative abundances of Li in nova explosions. The shaded area shows the range of observed values and the red horizontal dotted line is the average value.  Points connected by curves represent the highest values found in recent nova calculations; those of \cite{Jose2020} with 2-D models lie below the lower bound of the figure (adapted  from \cite{Borisov_2024b} and reference therein). 
}
\label{fig_Li_BBN_nova}
\end{figure}

\subsection{Explosive H-burning in novae}
\label{subsec_novae}

In one of the earliest attempts to explain the nova phenomenon,   a thermonuclear explosion of the H-layer at the surface of a white dwarf was invoked, triggered by 
$^3$He + $^3$He \citep{Schatzman_1951}. \cite{Gurevitch_1957} suggested that a plausible trigger of novae would be reactions involving the unstable isotopes of N ($^{13}$N) and O ($^{14}$O and $^{15}$O) 
with the overall rate of the CNO cycle  being determined by their beta-decay timescales, of the order of $\sim$100 s 
(the "beta limited hot CNO cycle").
This was confirmed quantitatively  with a hydrodynamic code incorporating a nuclear reaction network by \cite{Starrfield_1972}. 

The role of nova as potential producers of \lib \ was revealed in \cite{Arnould_1975} who performed  nucleosynthesis calculations of explosive H-burning with parametrized temperature and density profiles and noticed  the importance of the initial amount of $^3$He on the final \lib \ abundance. Their results were confirmed with the hydro+nuclear  code of \cite{Starrfield_1978}, who emphasized the impact of uncertainties of the nova modelling - in particular the speed and duration of convection - on the \lib \ production. Subsequent observations and calculations refined the picture of nova eruptions over the years \citep[see e.g.][for extensive reviews of nova properties and models]{Jose_2017}.
Regarding nucleosynthesis, there is general agreement that the main nucleosynthetic products are Li7, C13, N15, O17, all of them being formed with large overproduction factors of $\sim$10$^3$.  

For 40 years after the work of \cite{Arnould_1975} no observations of Li were reported in nova spectra. The line of Li I at  6708 \AA \ in the early high-resolution spectra of the “slow”
nova V1369 Cen was possibly detected by \cite{Izzo_2015}. The parent nucleus \bea \ was detected with high-resolution spectrographs in several recent novae and also found in reanalysis of archival data of historical novae from the {\it International Ultraviolet Explorer} \citep[][and references therein]{Molaro2023}. 

Nova observations show that Li/H  is on average almost 10000 times higher than the proto-solar one (Fig. \ref{fig_Li_BBN_nova}, right). This is in contrast with the predictions of almost all old and recent nova models \citep{Jose_1998, Starrfield_2016, Rukeya2017,Starrfield2024}, which produce at least 10 times lower values. 
Unlike previous studies, the one of \cite{Gao_2024} finds maximum \bea \ abundances  compatible with the observed ones, especially for CO novae. They attribute it to the high mass fraction of $^3$He assumed in their calculations combined to the treatment of element diffusion during the nova eruption.
However, 
\citet{Denissenkov2021}  found that a high abundance of $^3$He affects the energetics of the explosion (through the $^3$He+$^3$He reaction which ignites at lower temperature than $^{12}$C+p), decreases the peak temperatures and accreted masses and results in a reduced production of \bea. 

Nova calculations are still affected from several poorly known physical ingredients \citep{Jose_2017}.
Moreover, 1D models can hardly provide a realistic description of e.g. convection or the ignition of the explosion, which may occur in "hot-spots" and not simultaneously on the whole surface of the WD.  
Multidimensional hydro codes are definitely required for a more accurate description of the nova phenomenon. Despite the work of various groups for more than 30 years and the initially encouraging results regarding the core-envelope mixing, progress in that direction has been rather slow and contradictory results have been obtained sometimes  \citep[e.g.][and references therein]{Jose_2017}.
 Using a multiD model
 \citet{Jose2020} find results similar to those obtained with  1D models for most nuclear species, except for \bea \ for which considerably lower values are found.

Despite the disagreement between current nova models and observational findings on Li, observations support strongly the  role of novae as the main stellar source of Li. The observationally inferred Li yields of novae\footnote{The nova Li yields are derived from observations of Li/H taking into account the total mass of the ejecta, since H in nova has a considerably lower abundance than in the Sun.}, combined to the currently observed Galactic nova rates and despite the poorly known  nova history in the Galaxy, reproduce satisfactorily the protosolar Li abundance (see \S~\ref{subsec:Proto_solarLi_and_mainsource}).

\subsection{$\nu$ induced nucleosynthesis in core-sollapse supernovae}
\label{subsec_neutrinos}
In their seminal study on the hydrodynamical behaviour  of supernovae, \citet{Colgate_1966}  found that a) the large energy released by the gravitational collapse of the Fe core of a massive star ($\sim$10$^{53}$ ergs) would be mostly in the form of neutrinos emitted from the naissant proto-neutron star and that b) partial transfer of this energy to the stellar mantle could lead to a successful  explosion. 

\cite{Domogatsky_1977} suggested that in view of the huge number of the emitted neutrinos ($\sim$10$^{58}$) their interactions with the nuclei of the stellar envelope might have interesting nucleosynthetic effects, despite the extremely low values of the corresponding cross-sections which are $\sim$10$^{-45}$ cm$^2$.
Indeed, the energies of the neutrinos  reflect the temperature of the proto-neutron star (several $\sim$10$^{11}$ K) and are in the range of tens of MeV,  comparable to the excitation energies of nuclei. 
The excited nuclei decay through emission of photons, neutrons, protons, or $\alpha$ particles (and in some cases, also deuterons or $^3$He) and these particles interact in their turn with their hot and dense environment during the passage of the shock wave 
\citep{Domogatsky_1978}.
\cite{Woosley_1990} performed the first thorough investigation of the process, involving the evaluation of all relevant cross-sections and branching ratios and a simulation of the pre- and post-shock conditions in the various layers of a CCSN of 20 \ms. Regarding the light nuclei, they found that substantial amounts of \lib \ could be produced in the He-shell through $^4$He+$^3$He $\longrightarrow$ \bea \ and $^4$He+$^3$H $\longrightarrow$ \lib, with $^3$He and $^3$H produced by  de-excitation of $\nu$-excited $^4$He; moreover, significant production of  \bb \ could occur in the C-shell through de-excitation of $\nu$-excited $^{12}$C. 

Subsequent studies, with improved stellar, supernova, nuclear and  neutrino physics progressively refined that picture. 
Considering time-dependent spectra for all $\nu$ species from supernova simulations (rather than a parametric description of $\nu$-luminosity from the neutron star cooling phase)  and  the neutrino emission from the initial neutrino burst and accretion phases, \cite{Sieverding_2019} found that relevant amounts of \lib \ can be produced in a 27 \ms \ star (Fig. \ref{fig_Li_prod}, bottom panel).

 Neutrino-induced nucleosynthesis is a relatively recent and fascinating chapter regarding supernova explosions and the origin of the elements  \citep[see, e.g.][for a review]{Fischer_2024}. It offers a way  to explain the proto-solar \bba \ ratio, but most probably 
  it does not contribute to  the production of Li in the Galaxy(\S~ \ref{subsec:Li6Li7ratio}).

\section{LITHIUM AS A PROBE OF THE CHEMICAL AND DYNAMICAL EVOLUTION OF THE GALAXY} 
\label{sec:ChemicalEvolution}

The evolution of Li  was considered in  some of the earliest models  of Galactic Chemical Evolution (GCE thereafter) in the early 1970s \citep{Mitler_1970, Reeves_1973,  Truran_1971, Audouze_1974}. However, its fragility makes such studies difficult on two grounds: on the theoretical side, because of the uncertainties affecting the Li yields of the major candidate stellar sources (\S~ 4) and the time dependence of their rates; and on the observational side, because the  photospheric abundance of Li does not reflect, in general, the one at the star's formation (except for the hottest and youngest stars), thus preventing an unbiased comparison between model predictions and observations.

\subsection{The solar vicinity and the main Li source}
\label{subsec:Proto_solarLi_and_mainsource}

Most studies of Li evolution concern the solar vicinity (loosely defined as a cylinder of diameter $\sim$1 kpc perpendicular to the plane of the Galactic disk and centered on the Sun) and are made with simple 1-zone models. Such models produce a unique age-metallicity relation allowing for Fe to be used as a proxy for age, since stellar ages still suffer from considerable uncertainties.
Boundary conditions to such models are: a) the present-day Li abundance, b) the one at the Sun's formation (4.56 Gyr ago) c) the initial one (resulting from BBN) and d) the ‘upper envelope’ of the Li observations (hereafter ULiE).
\begin{marginnote}
\entry{Upper Lithium Envelope (ULiE)}{The highest Li abundance observed at a given metallicity, evaluated by taking the average  of (a few of) the most Li-rich stars in that metallicity range.}
\end{marginnote}

a) The Li abundance is  A(Li)=3.25$\pm$0.15 in the local ISM \citep{Knauth_2003}, and similar values were obtained for young stars in Orion \citep{Cunha_1995,Kos_2021} and in the young cluster NGC2664, estimated to be $<$5 Myr old \citep{Lim_2016}.

b) The proto-solar value, measured in meteorites, has been recently evaluated slightly upwards from previous ones, to A(Li)$_{\rm P-\odot}$=3.39$\pm$0.02 \citep{Lodders_2025}.

c) SBBN gives values of A(Li)$_{\rm SBBN}\sim$2.70 (\S~ \ref{subsec_bigbang}), implying that the primordial contribution to the proto-solar Li abundance is $\sim$20\%. A similar contribution comes from GCR (\S~ \ref{subsec_CosmicRays}).

d) In the past the ULiE has been interpreted  
as a tracer of the true Li evolution \citep{Rebolo_1988}, and it was used  - incorrectly - to determine the timescale of the stellar Li source. However, the leading alternative interpretation  \citep{1991MNRAS.253..610L} is that ``{\it the galactic Li abundance at a given [Fe/H]
exceeds the observed maximum abundance and has been reduced in all stars to give the observed abundances}", i.e. the  ULiE constitutes a lower limit constraining the true Li evolution from below.

The  role of the various candidate Li sources has been scrutinized over the years  \citep{Audouze_1983,DAntona_1991, Abia1995, Matteucci_1995,Romano2001,Travaglio2001,Prantzos2012}. Available Li yields of all stellar sources seemed insufficient to  make the required amount of Li. 
In the past ten years novae attracted considerable attention, due to the detection of substantial abundances of \bea \ in their ejecta (\S~ \ref{subsec_novae}).
One-zone GCE models   using Li yields inferred from observed nova abundances now reproduce the meteoritic Li value \citep{Rukeya2017,Cescutti2019,Grisoni2019, Romano2021,Kemp_2022b, Gao_2024, Borisov_2024b, Nguyen2025}. These models make different assumptions on the history of the nova rate, which is only constrained from its observed current Galactic value of $\sim$30-50 yr$^{-1}$ \citep{Shafter_2017,De_2021,Kawash2022}
It constitutes a critical ingredient of the models,  non-trivially connected to the assumed star formation rate through the nova Delayed Time Distribution (DTD). 
\begin{marginnote}
\entry{Delayed Time Distribution(DTD)}{ The frequency distribution in time of some stellar events (AGBs,novae, SNIa) after the formation episode of their progenitor sources
}
\end{marginnote}

Various evaluations/approximations of the nova DTD have been  proposed in the literature and adopted in the GCE models. The most complete study is the population synthesis  of \citet{Kemp_2022a}  accounting for various factors on the evolution of binary systems (mass, initial separation and metallicity), which -interestingly -  {\it predicts typical ejecta masses compatible with observed ones} and a strong  dependence of the nova DTD on the metallicity of the progenitor system.  
Besides the metallicity dependence, a peculiar feature of that DTD (a power law as function of time) is that its slope $X_{\rm NOV}$ is systematically shallower -  for all metallicities - than the corresponding one of SNIa ($X_{\rm SNIa}\sim$--1). The production ratio Li(novae)/Fe(SNIa) increases  with time after a burst of star formation and particularly at late times\footnote{In  the case of SNIa,  observations of external galaxies are used to evaluate reasonably well their DTD$_{\rm SNIa} \propto$ time$^{-1}$ but no such observations exist for novae. The production ratio  Li(novae)/Fe(SNIa) varies with time $t$ as $t^{X_{\rm NOV}}$/$t^{-1}$=$t^{(X_{\rm NOV}+1)}$, i.e. it increases with time for all values of $X_{\rm NOV}>-1$, which is the case of the DTDs of \cite{Kemp_2022b}. }, with the characteristic timescales being of a several  10$^8$ yr for SNIa and $>$1 Gyr for novae. 

As a result,  {\it for the same metallicity} a slowly evolving system reaches a higher Li abundance than a rapidly evolving one, because in the latter case, metallicity increases faster, in such short timescales that most long-lived sources - like novae - have no time to release their Li 
\citep{Prantzos2017}. The DTD of Li sources  has important implications for the interpretation of the Li observations, both locally and Galaxy wide. A short-timescale  source of Li breaks at earlier times and lower metallicities the primordial Li plateau than a long-lived one, but it leads to a smaller Li increase at late times and  high metallicities. This feature explains most of the differences between GCE models of Li having novae as the main/only stellar source.

It should be emphasized that 1-zone models like the one in Fig. \ref{fig_Li_evol_Mz} (left panel), represent poorly the situation, either  in the Galactic halo (\feh$<$-1), or in the local disk. Indeed, the local super-solar metallicity stars were suggested to originate from the inner disk \citep{Grenon_1989}, and the large scatter in the local age-metallicity relation from diffusion of stars from other regions to the solar neighborhood \citep{Edvardsson_1993}; we discuss those points, crucial for understanding Li evolution, in the next section. On the other hand, the halo is thought to be formed by an ensemble of smaller  subhaloes with different histories and age-metallicity relations. However, the abundance ratios between primary elements having the same source (e.g. O and Mg, as well as Fe from CCSN) are expected to be constant as a function of time. The same holds for the GCR components of Li \ (\lia \ and \lib) calculated as described in \cite{Prantzos2012}, in which case the evolution depicted in Fig. \ref{fig_Li_evol_Mz} (left) for the contribution of GCR can be considered  ``realistic" (see \S~ \ref{subsec:Li6Li7ratio} for  alternatives).
\begin{marginnote}
\entry{Primary elements}{Elements with yields independent of the initial metal content of their parent star, like e.g. O or Fe from massive stars.}
\end{marginnote}

\begin{figure}[h]
\begin{minipage}{.55\textwidth}
{\includegraphics[width=\textwidth]{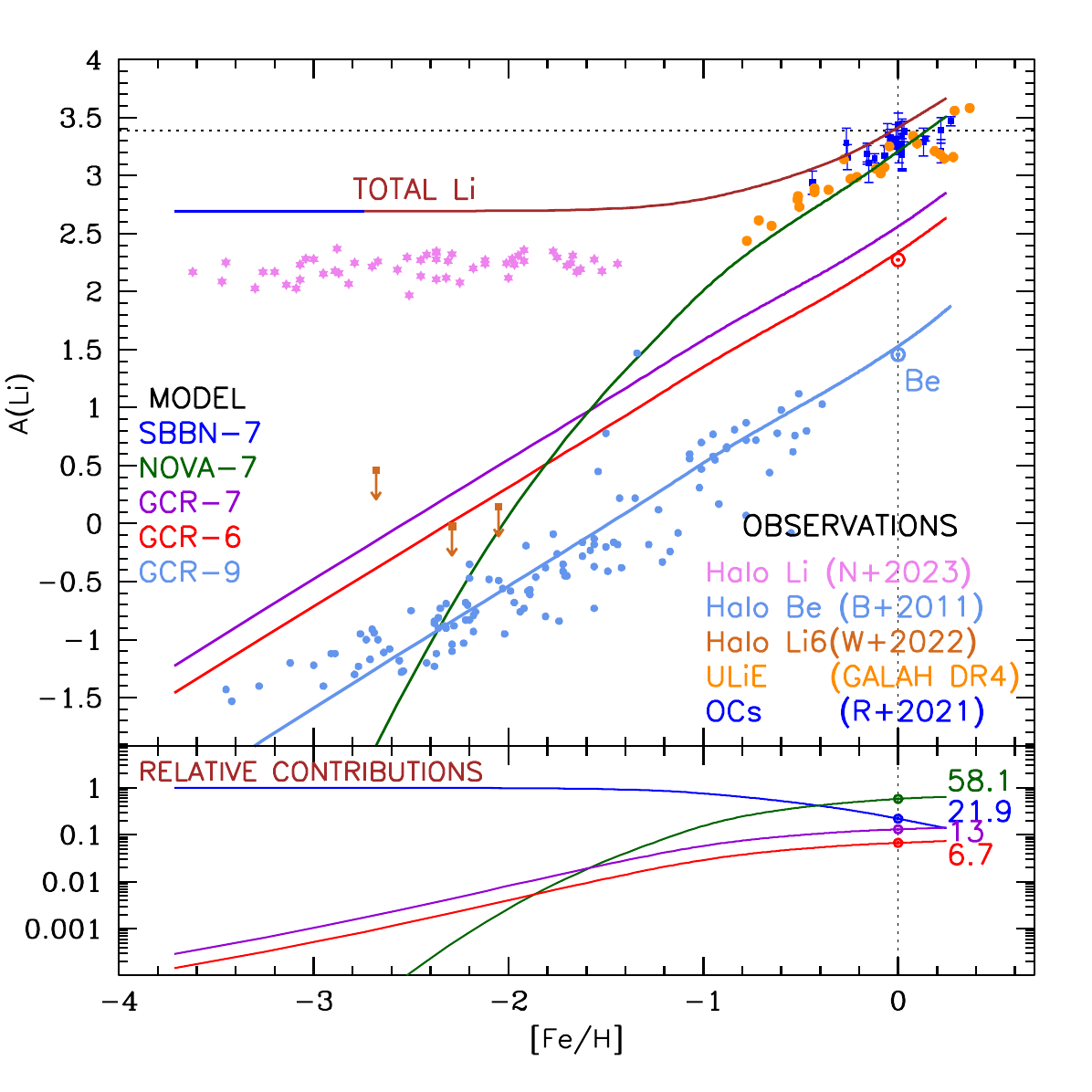}}
\end{minipage}
\hfill    
\begin{minipage}{.55\textwidth}
{\includegraphics[width=\textwidth]{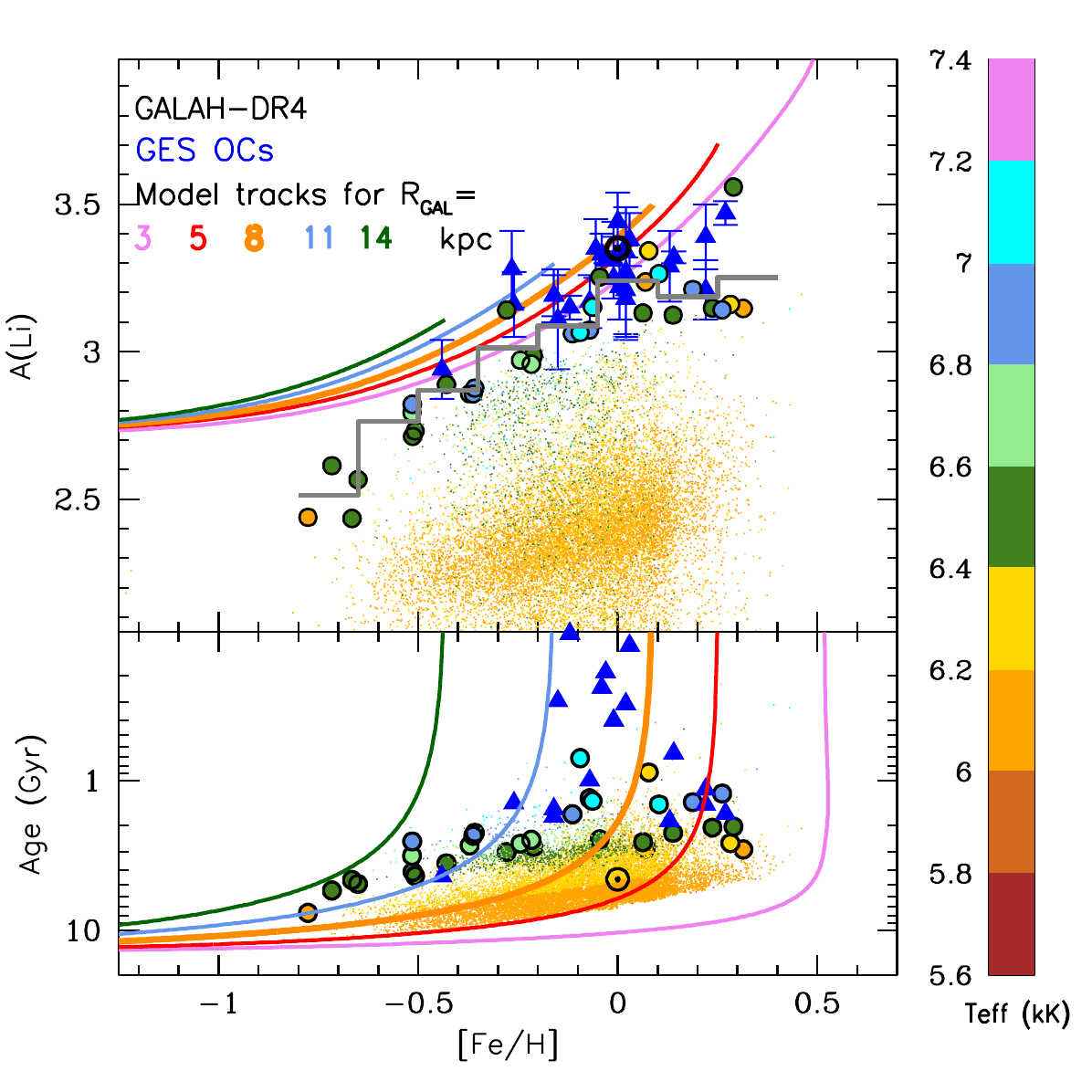}}
\end{minipage}

\caption{
{\it Left top:} Evolution of Li and its components (BBN and novae for \lib \  and GCR for \lia \ and \lib) and of \beb (a ``monitor" of GCR activity and composition)  \ as a function of [Fe/H] in a simple 1-zone model of the solar neighborhood ({\it top}). Data from \citet[][for Li]{Norris_2023} with corrections applied in \citet{Borisov2024}, \citet[][for Be]{Boesgaard_2011}, \citet[][for Li-6 upper limits with downward arrows]{Wang_2022},
\citet[][for Open Clusters]{Romano2021} and \citet[][GALAH DR4 Upper Li Envelope ULiE]{Buder2025}.\\
{\it Left bottom:} Contributions of individual components,  with numbers indicating the corresponding percentage contributions at Sun's formation. \\
{\it Right top}:  a zoom on the upper right part of the top left figure, with field star data from GALAH DR4  in the solar neighborhood (7.5$<$R$_{G}$/kpc $<$8.5), color coded for temperature with filled circles indicating the 4 most Li-rich stars in each [Fe/H] bin and the grey histogram indicating their average trend; blue triangles with error bars indicate OC data  (no colour coding for temperature), while tracks of gas abundances of Li vs Fe in various galactocentric radii are from a multi-zone model of the Galaxy \citep{Borisov_2024b}. \\
{\it Right bottom}: Corresponding stellar ages vs \feh \ with same colour-coding as in the top for the data and the model tracks.
}
\label{fig_Li_evol_Mz}
\end{figure}

\subsection{Evolution of Li in the Milky Way thin disk}
\label{subsec:Li_inMWdisks_RadMigr}

An early attempt to extend the study of Li evolution from the solar vicinity to the whole Galactic disk was made in \cite{Prantzos2012} with an independent-ring  multi-zone model; adopting a long-lived stellar Li source, he predicted a present-day Li gradient of d[Li/H]/dR= -0.05 dex/kpc, not very different from the one of oxygen.
However, \cite{Sellwood_2002} revealed the importance  of radial migration of stars for the evolution of galactic disks.
Radial migration turns out to be crucial in interpreting Li observations in the Milky Way, even more so than for other, less fragile elements.

A first ``alert" in that respect came from the unexpected finding of a non-monotonic behaviour of the ULiE, namely a decline of Li/H at super-solar metallicities  \citep{DelgadoM_2015,Guiglion_2016, Fu_2018, Bensby_2018}, with different data samples
\footnote{\citet{Prantzos2017} explored the idea of strong reduction of AGB yields of Li at such high metallicities, while acknowledging that it is not supported by the stellar nucleosynthesis models of \citet[][see \S~ \ref{subsec:HBB_AGB}]{Karakas2016}; they suggested that strong Li depletion in the stellar envelopes  may contribute to, or even be at the origin of, the effect.}. 
The extensive investigation of \citet{DelgadoM_2015} revealed that other properties of the ULiE also vary in a non monotonic way with metallicity, going through an extremum at \feh$\sim$0 (age, mass, average T$_{\rm eff}$).
\citet{Guiglion2019} suggested that during the several Gyr time of their migration from the inner disk to the solar vicinity, super-solar metallicity stars may have depleted their original (presumably super-solar) Li to  the presently observed levels. Their suggestion was confirmed by the investigation of \citet{Minchev_2019} who used mono-age populations from the HARPS survey and \citet{Dantas_2022} who included stellar kinematics from GES data.

Several surveys in the past decade have increased considerably the size and the quality of the data samples regarding Li in the Galactic disks: HARPS \citep{DelgadoM_2015}, AMBRE \citep{Guiglion_2016}, LAMOST \citep{Gao_2019}, GES \citep{Randich2020,Romano2021}, GALAH DR3 \citep{Gao2020, Wang_2024} and DR4 \citep{Buder2025}. 
The trend of declining Li/H was not confirmed by observations of stars hotter than the ``Li-dip"in  GES data of Open Clusters \citep[OCs,][]{Randich2020} and similar conclusions  were reached by the analysis of GALAH DR3 data   by \citet{Charbonnel2021} who discussed the potential impact of atomic diffusion on the Li abundances of those stars.

 The aforementioned results are summarized in Fig. \ref{fig_Li_evol_Mz} (right) in terms of stellar temperatures and ages, using data from GES OCs and the GALAH DR4 field stars in the solar vicinity \citep{Buder2025}\footnote{The GALAH DR4 stars used in Fig. \ref{fig_Li_evol_Mz} have logg$>3.8$, signal-to-noise ratio SNR$>$80, Vsin$i<$20 km/s. NLTE corrections  \citep[as found in][for GALAH DR3 stars]{Wang_2024} have not been applied, but for hot stars they are insignificant (X. Wang, private communication).}. At each metallicity bin, the ULiE is mainly formed  by the hottest (top panel) and youngest (bottom panel) objects of that bin, the trend being essentially  monotonic with metallicity, except for the highest metallicities where it is reversed. The GALAH data concern a local stellar population, within less than a kpc from the Sun, but the large dispersion in their age-metallicity relation  is a clear signature of stellar radial migration (bottom panel in Fig. \ref{fig_Li_evol_Mz}, right, and Fig.~1S in Supplemental Material of the electronic edition).

The combined effects of stellar Li-depletion and radial migration render   the use of \feh \ as a proxy for time obviously futile, even for stars found locally today. Therefore, the ULiE cannot be used to determine the timescale of the main stellar Li source as often done  in the past. The locally observed  ULiE is composed of stars formed in different places  with different star formation histories and it is  modulated by age and metallicity dependent Li depletion  and  radial migration (which determines the fraction of stars brought to the local volume from other radii). 
The timescale of radial migration ($\sim$1 kpc/Gyr), makes it statistically improbable - albeit not impossible - to find  very young ($<$1 Gyr) and hot  ($>$6800 K) stars of highly super-solar metallicity and very high Li values in the solar vicinity. The chances of finding such stars should increase with the size of the sample.


Similar results with those displayed in  Fig. \ref{fig_Li_evol_Mz} (right), at least qualitatively, are obtained with the field stars of other surveys, like AMBRE \citep{Guiglion_2016} or GES \citep{Romano2021}. 
Moreover,  GES provides data for OCs  in a range of Galactocentric radii 4 $<$R$_G$(kpc)$<$14, thus offering important supplementary information.
However, 
the effects of radial migration make it difficult to infer the present Li gradient from observations and to compare with models \citep[see discussions in][]{Romano2021,Borisov_2024b}.

Although the importance of radial migration in interpreting the Li observations has been properly emphasized in \citet{Minchev_2019}, a quantitative understanding of the impact of this effect  is still lacking because of the difficulty of `unmixing" the stellar populations of the disk. Attempts to determine the {\it birth radii} of stars  for Li studies  were made recently  \citep{Minchev_2019,Zhang2023,SunT_2025,Dantas_2025}, based on various stellar samples and methods. Methods dependent on GCE models and augmented by analysis of kinematic and dynamical data are recently developed in \citet{Dantas2025a}. Methods independent of the GCE models, inferring  the evolution of the Galactic metallicity gradient at birth radius from observations  \citep{LuY_2024, Ratcliffe_2023, Ratcliffe_2024} are promising but not yet free from biases and rely on critical assumptions \citep{LuY_2022, Chen_2025}. The application of a simple method, suggested in \citet{Minchev2018}, is illustrated for the case of Li in Fig. 2S of the Supplemental material.

A quantitative assessment of the Li evolution in the Milky Way disk will  require more elaborate, model independent,  methods  to determine the evolution of the metallicity gradient at birth place of stars. Such methods will need large and unbiased stellar samples with accurate age determinations,  as well as dynamical and kinematic information from observations.
Combining the chemodynamical evolution of the Galaxy to the study of Li in stars of various types, ages, metallicities, masses and rotational velocities will constitute a powerful probe of
galactic and stellar evolution: the stellar age could be compared to the timescale of radial migration  from its birthplace to the solar vicinity and to the timescale for its Li depletion expected from stellar models. It could thus constrain both those stellar and galactic physical effects, 
opening  a new  window for the study.

Despite the large number of stars with Li detection in current surveys, a much larger number of super-solar metallicity stars from the warm side of the Li dip will be necessary to assess the Li evolution in the \feh$>$0.1 region. Such large numbers will be needed for the training samples of machine learning techniques that could be used in the data analysis of next generation spectroscopic surveys. The potential of Convolution Neural Networks to exploit large sets of Li data is illustrated in \citet{Nepal2023}. 

\subsection{The evolution of Li in the thick disk and the bulge }
\label{subsec_ThinThickBulge}

Despite the large number of studies over the years, there is no clear identification of the thin and thick disks of the Galaxy, since their various properties (morphological, kinematic, chemical, stellar ages) overlap to various extents, the overlap depending on the galactic location \citep[see e.g. ][]{Imig_2023}. Thus,  the role played by the various processes  invoked to explain the differences between those properties still remains unclear. An example is the observed double-branch behaviour of the \afe \  ratio in the solar vicinity: high \afe \ for the older thick disk and low \afe \ for the younger thin disk at a given metallicity.

The behaviour of Li in the thick disk was investigated observationally  in several studies \citep{Molaro_1997, Romano_1999, Ramirez_2012, DelgadoM_2015,
Guiglion_2016, Bensby_2018, Fu_2018} who adopted different criteria for the definition of the thick disk (chemistry, kinematics, or age) and found different trends of Li/H vs \feh,  namely declining, increasing or constant.

The evolution of Li in the ``low \afe" and ``high \afe" disks was studied up to now  with semi-analytical models, which can be classified in two categories:
One class  invokes  models without radial migration, with two regimes of star formation in the MW (a fast and a slower one, occurring either sequentially or in parallel). They are separated by a pause in star formation during which the gaseous abundances at the end of the first regime are diluted by the infall of pristine material \citep{Grisoni2019, Cescutti2019, Romano2021}. 
The second class concerns multi-zone
models with radial migration \citep{Prantzos2017,Borisov_2024b}%
The ``high \afe" branch of the local population
results from the presence of old ($>$9 Gyr) migrant stars formed early on in the inner Galaxy while the ``low \afe" branch is composed from stars formed locally or are migrants from  the outer disk  (at subsolar metallicities) or from the inner disk (at super-solar metallicities). 

In all cases,  the model evolution of Li in the gas of the high \afe \ disk is determined mainly by the adopted pre-galactic abundance (SBBN vs Halo Plateau) and the timescale of the adopted Li stellar source. High pre-galactic abundances and/or long timescale sources produce a slow (or null) increase of Li in the gas of the high \afe \ phase of the Galaxy \citep{Prantzos2017, Cescutti2019, Grisoni2019}. This is not necessarily reflected in the observed stellar Li abundances of the high \afe \  disk  which are affected by depletion to various extents and may lead to declining trends of Li/H vs metallicity \citep[see][using monoage samples from AMBRE data]{Minchev2018}. 
An understanding of the impact of the various stellar parameters  
on the amount of that depletion is required in order to decipher the early and late evolution of Li, as attempted in \citet{Nguyen2025}.

Similar conclusions hold for the evolution of Li in the Galactic bulge, explored observationally by \citet{Bensby_2020}, who found a decline of Li/H at low metallicity,  akin to the one of the thick disk. The existence of several stellar components in the bulge \citep{Wylie2021} suggests 
a complex history for that region, involving stellar populations with different star formation histories. The interpretation  of the Li abundances in the bulge would be at least as complex as the one for the solar neighborhood. 
A non-monotonic behaviour of Li/H with age or metallicity is expected, modulated by depletion.

\subsection{The  evolution of the \lir \ ratio }
\label{subsec:Li6Li7ratio}

Historically, the \lir \ ratio played an important role in studies of the light element evolution, because the lighter isotope has only one - and well understood - source, namely GCR. Thus, any important variation of that ratio between its meteoritic value and the one in the ISM  would probe the differential activity between GCR and any stellar source of \lib \  in the past 4.5 Gyr in the Galaxy \citep{Reeves_1993}. However, the situation becomes more complicated by the role of several factors, such as  recent infall, presumably of primordial composition, which should dilute more \lia \ than \lib \ (since the former does not exist in primordial matter). Measurements of the  \lir \ in the local ISM \citep{Kawanomoto_2009, Knauth_2017} provide  values compatible with the meteoritic one and with simple GCE models \citep{Prantzos2012}. On the other hand, significantly  higher values of \lir \ (factors 2-4) have been reported in several lines of sight to Galactic regions of stellar or supernova activity, like the SN remnant IC433 \citep{Taylor_2012} or the star forming region IC 348 \citep{Knauth_2017}. This clearly argues for intense  activity of energetic particles accelerated by local supernovae and disfavours any contribution of the $\nu$-process to \lib. In contrast, the upper limit of \lir $<$0.1 obtained in the line of sight towards Sk 143 in the LMC \citep{Molaro_2024} is compatible with the meteoritic value.

The behaviour of \lia \ in halo stars was extensively investigated because it may provide information either about non-standard primordial nucleosynthesis or about the physics of GCR - spectra, energetics, composition, confinement - in the early Galaxy ,\citep{Prantzos_1993, Fields_1999,Ramaty_2000, Prantzos_2006, Prantzos2012}. Up to now, only upper limits on \lia  \ have been obtained \citep{Lind_2013,Wang_2022}. On the theoretical side, \lia \ was expected to behave as a primary element at early times, its production been
dominated by $\alpha+\alpha$ fusion reactions \citep{Steigman_1992}. 
\begin{marginnote}
\entry{Secondary elements}{Their yields depend on the metallicity of their parent star (e.g. s- elements from AGB stars) 
}
\end{marginnote}
The underlying assumption is that GCR are accelerated from the ISM and therefore CNO spallation reactions produce always secondary Li, Be and B. However, observations of  halo stars in the 1990s found an unexpected primary behaviour of Be (as depicted in Fig. \ref{fig_Li_evol_Mz}, left). \citet{Prantzos2012} suggested that GCR are accelerated in the winds of {\it rotating massive stars} always expelling primary CNO, in which case it is the spallation reactions of CNO that dominate the production of \lia \ throughout the galactic history and not $\alpha+\alpha$ fusion \citep[see, however][ for a different viewpoint on the acceleration site of GCR]{Tatischeff_2021}. In
this oversimplified picture of LiBeB evolution, 
\lia \  is expected to evolve in line with Be  as primary, and its evolution is compatible with the current upper limits (see Fig. \ref{fig_Li_evol_Mz}, top left), particularly if its nuclear fragility is taken into account. In contrast, using different assumptions about LiBeB production (a peculiar evolution of O/Fe and ``overconfinement" 
of GCR in the early Galaxy) \citet{Fields2022} find a considerably higher abundance of \lia, which
evolves as \feh$^{0.5}$  (a kind of ``overprimary").
They argue then that such a high theoretical abundance, compared to the current upper limits,
implies a large destruction of \lia \ and consequently of \lib, independently of any models of Li destruction in stars. However, \lia \ is destroyed $\sim$100 times faster than \lib \ (Fig. \ref{fig_Nucleardestruction-abundanceprofiles}, left)
and its destruction should not constrain significantly the one of its heavier isotope.

In any case, the monitoring of the \lir \ ratio in early or late galactic environments can provide extremely useful information on various physical processes, in particular regarding the cosmic ray activity in comparison to the one of the stellar source of \lib.

\begin{summary}[SUMMARY POINTS]
\begin{enumerate} 
\item Compared to seismic solar-analogue stars, the Sun has a normal photospheric Li abundance and rotation rate at its present age. 
\item The Li dip is a universal main sequence Li depletion phenomenon  confined to the same effective temperature range for all metallicities at which it is potentially observable (i.e., down to [Fe/H]$\sim -1$). All the evidence suggests that rotation-induced mixing is responsible for this pattern when stellar winds efficiently extract angular momentum early on the main sequence.
\item The universality of the Pop~II Li plateau for [Fe/H] lower than $\sim -1.0$
in the Milky Way and other galaxies 
implies that this feature does not result from the astration of a large fraction of the interstellar medium (ISM) of galactic haloes before the formation of Pop~II stars. Instead,
the discrepancy between the Li value predicted by BBN+CMB
and the one observed along the Spite plateau has a stellar solution. 
\item Type~II rotating models including the transport of angular momentum and chemicals by meridional circulation and shear turbulence with specific prescriptions, together with a simple modelling of internal gravity waves, can simultaneously account for the Li abundance patterns in Pop~I and II dwarf stars, including the solar Li, the Li dip, and the Spite plateau. 
\item Thermohaline instability is currently favoured as the mechanism that can consistently explain the drops in lithium abundance,  carbon isotopic ratio, and C/N ratio at the luminosity of the RGB bump. \item When accounting for realistic stellar mass distribution in samples of red giants observed by large spectroscopic surveys together 
with a proper mass-dependent Li depletion on the MS, there is no evidence  of an unknown Li production mechanism occurring between the upper RGB and the clump or the horizontal branch. 
\item  The proto-solar Li - meteoritic value A(Li)=3.39  results from a mixture of nucleosynthesis products of  3 different sites: early universe ($\sim$20\%), Galactic Cosmic rays ($\sim$20\%) and stars ($\sim$60\%). On observational grounds, novae are currently favoured as the main/only stellar source.
\item  The evolutionary history of Li is poorly understood theoretically (in view of uncertainties in the Delay Time Distribution of novae) and hard to constrain observationally (because of depletion effects, depending on various stellar properties). The ULiE does not represent the evolutionary history of Li, it only constrains it providing lower values.
\item Stellar radial migration is a key to decipher the ULiE and, more generally,  the observations of Li in the local volume and the disk of the Galaxy. Most of the  super-solar metallicity stars observed locally are ``messengers" from the inner disk and sub-solar metallicity ones from the outer disk, i.e., they probe regions with different star formation and Li histories.
\end{enumerate}
\end{summary}

\begin{issues}[FUTURE ISSUES]
\begin{enumerate}
\item The Li-age-mass-\teff-[Fe/H]  relations among 
different populations of stars derived from present and future surveys 
must be meticulously scrutinised through the modelling of individual frequencies (to get the stellar mass, radius, age), combined with exquisite spectroscopic (to get the stellar effective temperature, metallicity, Li), photometric (to get P$_{rot}$), and astrometric (to get the stellar luminosity) data, together a with careful examination of the bias of the observed samples.  
\item Li abundance in extremely metal-poor dwarfs with \teff ~above $\sim$ 6000~K and no significant C-enrichment should lie on the Li plateau. Although such objects should be rare, they are worth searching for.
\item The measurement of the rotation of the deep solar core below 0.2~R$_{\odot}$ is required to legitimate the origin of the angular momentum transport process in the Sun and low-mass stars and its impact on the transport of chemicals that drives Li depletion. 
\item The major limitations of the stellar evolution models come from the lack of understanding on shear turbulence, magnetic and hydrodynamic instabilities, and their interactions with (magneto)-gravito-inertial waves. There is an imperative need for extended formalisms that can self-consistently account for these mechanisms in stellar evolution models. 
\item Important asteroseismic efforts must be devoted to ascertain the mass and the position in the HRD diagram of super Li-rich giant candidates.
\item A better understanding of Li production in nova (with  extra ingredients or full 3D models) will be required  to firmly establish their importance as main stellar Li producers. The same holds for the nova DTD.
\item Radial migration opens a new chapter in Li studies, which should rather be made on a star-by-star basis,
combining the stellar properties and corresponding stellar depletion models with Galactic properties and chemo-
dynamical models. The two  aspects, stellar and galactic, are linked by the stellar ages for which a  more accurate evaluation is required.
\item The field will benefit from  next generation spectroscopic surveys (4MOST,
WEAVE, 4MIDABLE). A much larger number of super-solar metallicity stars from the warm side of the Li dip  will be necessary to explore the Li evolution in that extreme metallicity region.
\end{enumerate}
\end{issues}

\section*{DISCLOSURE STATEMENT}
The authors are not aware of any affiliations, memberships, funding, or financial holdings that
might be perceived as affecting the objectivity of this review. 

\section*{ACKNOWLEDGMENTS}
We are grateful to our colleagues who read and commented different sections of the manuscript, namely, 
L. Amard, S. Borisov, K. Cunha, J. José, N. Lagarde, I. Minchev, P. Molaro, A. Palacios, O. Richard, and S. Vauclair. We thank 
E. Wang for advice on GALAH DR4 Li data, and S. Randich and E. Franciosini for advice on GES data. We are grateful to S. Borisov for his kind help with the data management.
CC acknowledges support from the Swiss National Science Foundation (SNF; Project 200021-212160, PI CC) and IAP for their hospitality during the writing of this review.

\section*{SUPPLEMENTAL MATERIAL to Sec. 5.2: Evolution of Li in the Milky Way thin disk. 
{\bf Included only in the electronic edition of ARAA}}

The same models and GALAH data as in Fig. 8 (right) of the Main Text (hereafter MT) are plotted as a function of age in Fig.~9 of the Supplemental Material (hereafter SM). Independently of the model tracks, observations suggest that the 
stars of the ULiE (the most Li-rich stars in the metallicity bins of Fig. 8) trace also an upper envelope for a given age (bottom panel of Fig.~9). However, the  subsolar metallicity stars of the ULiE are, rather counterintuitively,  systematically among the most metal-poor ones at a given age  (see top panel of Fig.9). The reason is that metal-poor stars have lower opacities, are  hotter and have less extended envelopes than metal-rich stars of the same age; as a result, they deplete less their Li than their cooler counterparts. Indeed, a gradient of T$_{\rm eff}$ from cooler to hotter stars is observed in the  metallicity vs age panel (top panel in Fig.1) but an inverse gradient, from hotter to cooler stars, is observed in the lithium abundance vs age panel (bottom). 
Thus, at a given age, the hottest stars of sub-solar metallicities  are the most metal-poor ones and preserve better their Li. 
In view of the observed negative Galactic metallicity gradient, those stars which compose the sub-solar metallicity ULiE, are logically formed in the outer disk. 
{\it Thus, observations of \feh \ and Li vs age in local stars, combined to our understanding of the metallicity impact on the physics of stellar envelopes,  provide another strong argument for radial migration.}

\begin{figure}[h]
{\includegraphics[width=0.95\textwidth]{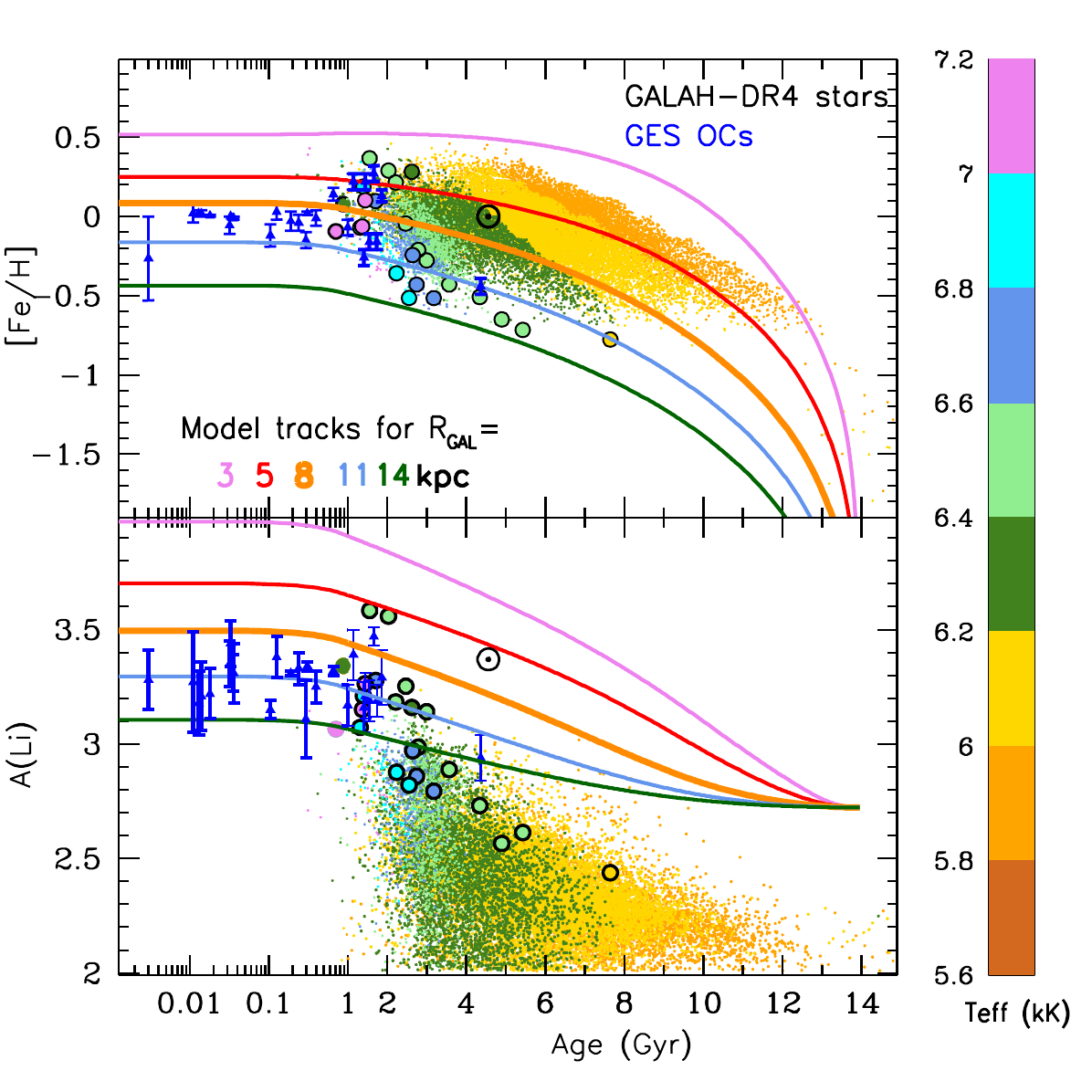}}
\caption{{\it } 
Same data, model tracks and color coding as in the right panels of Fig. 8, for GALAH DR4 local stars and GES OCs, plotted as function of age for \feh \ (top) and Li (bottom); the sub-solar metallicity stars of the ULiE  are, in general, the metal-poorest ones (top panel) in each age bin, except for the case of the youngest ones (1-2 Gyr) which may have both sub-solar and super-solar metallicities..}
\label{fig_Li_age}
\end{figure}

This confirms the conclusions of Sec. 5.2 (MT), namely that  abundance patterns of Li observed in the solar vicinity result from a complex interplay between stellar ages, metallicities and radial displacements (with other factors, like mass and rotation also playing a role).  The ULiE for [Fe/H] $<$-0.1 is composed of stars aged from a few to several Gyr from the outer disk brought in the solar vicinity by radial migration. Stars older than $\sim$8-9 Gyr and even lower [Fe/H] were formed instead in the inner disk, since the inside-out formation of the disk produces very few stars in the solar vicinity  at such early times. 
As for the super-solar metallicity stars of GALAH and OCs (less than 2 Gyr old) which are presumably formed in the inner disk (see Sec. 5.2 in MT), it is at present unknown whether they suffered some Li depletion and by how much, because the theoretical predictions depend on the DTD of the main stellar source.

\citet{Minchev_2019} were the first to emphasize that ``{\it ... the study of the production/destruction of lithium as  function of birth radius will provide stronger constraints on chemical evolution models”}, not only regarding the super-solar but also the sub-solar metallicity regime. However, the effects of radial migration make it difficult to infer the present Li gradient from observations and to compare with models.
Attempts to determine the {\it birth radii} of stars  for Li studies  were made recently :  \citet{Zhang2023}
evaluated birth radii R$_B$ of stars using the simple prescription of \citet{Minchev2018} for the evolution of the Galactic gas metallicity gradient over time. \citet{SunT_2025} used GALAH DR3  data combined to ages derived from {\it Gaia} and adopted the method of \citet{LuY_2024} to derive the birthplaces of the sample stars, while \citet{Dantas_2025}   derive R$_B$ by comparing positions of observed stars in the (R$_G$,\feh) plane  to results of the GCE model of \citet{Magrini2009}, a model-dependent method.  

The application of a simple method, suggested in \citet{Minchev2018}, is illustrated in 
Fig. \ref{fig_Li_evol_vs_age_radius} for the GALAH DR4 stars of the ULiE: stars observed locally originate from the inner or the outer disk, in birth places depending on their age and metallicity (upper panel);  their Li abundance today is lower than the one predicted from the GCE model at their inferred birth radius and the time of their birth (lower panel). This figure merely illustrates the principle and the potential of the method, since the results  obviously depend on the  adopted GCE model (including the  nova Delayed Time Distribution) and on the selection biases  affecting the ULiE of the  GALAH DR4 sample.

\begin{figure}[h]
{\includegraphics[width=0.95\textwidth]{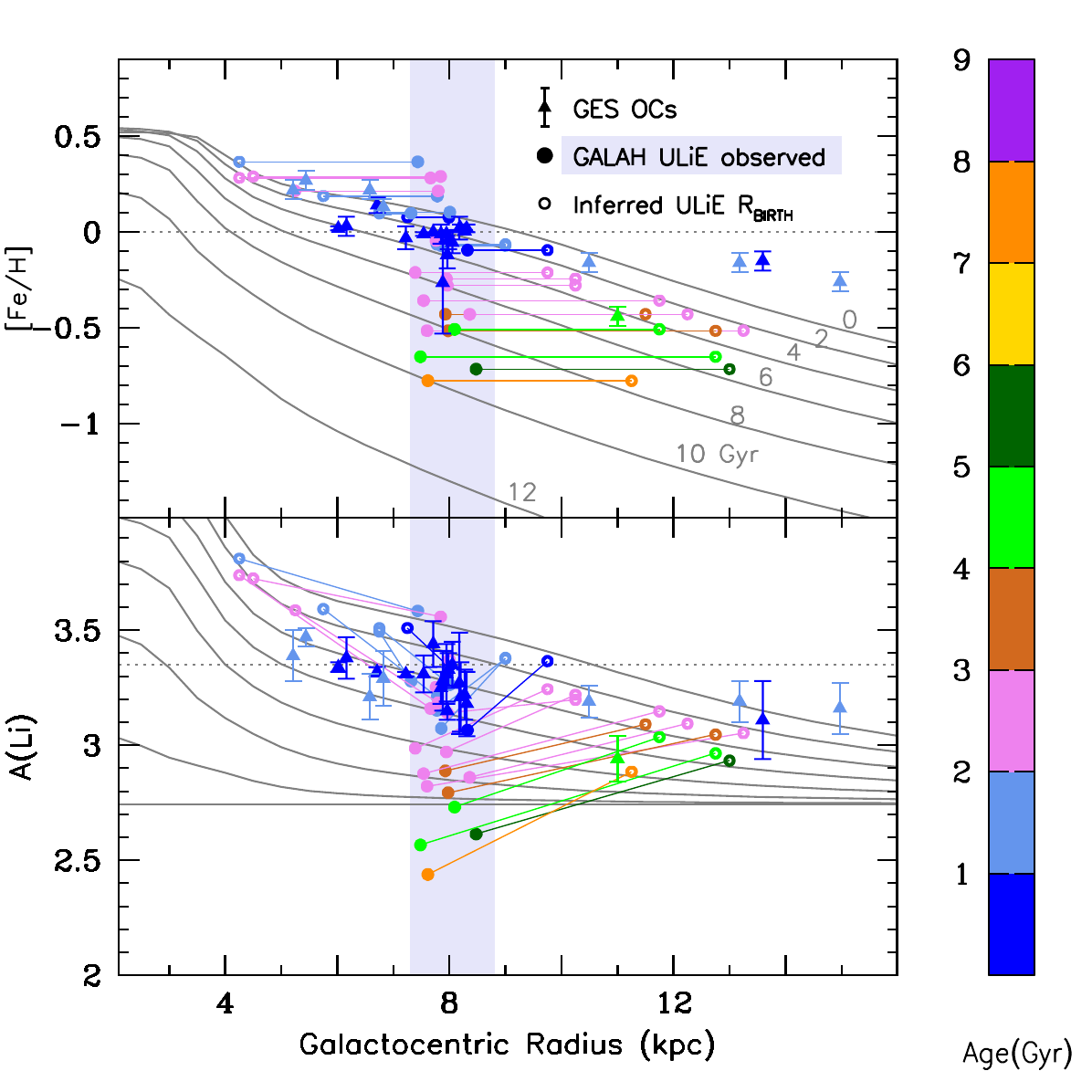}}
\caption{GES OC data \citep[][triangles with error bars]{Romano2021} vs Galactocentric radius. Field stars of the GALAH DR4 ULiE, observed in the region of R$_{\rm GAL}$= 7.5-8.5 kpc, are plotted within the shaded aerea (filled circles). OCs and field stars are colour-coded with age. The stars - but not the OCs - are "projected" (top panel) into their birth radius (open symbols connected to the filled circles) by {\it assuming} that the \feh-metallicity relation in the Galaxy is well reproduced at every radius by the GCE model of \citet{Prantzos2023}, as represented by the iso-age curves (in grey, labeled with the corresponding ages). The Li abundances of the model \citep[calculated in ][]{Borisov_2024b} at the inferred stellar birth places of the top panel are then connected to the locally observed ones in the  bottom panel. While the Fe abundance is assumed to suffer no depletion,  Li is obviously depleted during  migration to the local volume, either from the inner or the outer disk.
}
\label{fig_Li_evol_vs_age_radius}
\end{figure}

%

\newpage

\bibliographystyle{ar-style2}

\bibliography{Lithium1}






\end{document}